\documentclass[12pt,letterpaper]{article}
\usepackage[utf8]{inputenc}

\usepackage{graphicx,array}
\usepackage{url}
\usepackage{color}
\usepackage{latexsym}
\usepackage{amsthm}
\usepackage{amsmath}
\usepackage{amssymb}
\usepackage{amsfonts}
\usepackage[numbers,sort&compress]{natbib}
\usepackage{bm}
\usepackage{bbm}
\usepackage{slashed}
\usepackage{mathrsfs}
\usepackage{enumerate}
\usepackage{tikz}
\usepackage{siunitx}
\usepackage{mdframed}
\usepackage{setspace}  
\usepackage{esvect}
\usepackage{physics}
\usepackage{enumitem}
\usepackage{booktabs}
\usepackage{tcolorbox}%

\numberwithin{equation}{section}

\usepackage{hyperref} 
\hypersetup{
    colorlinks=true,       
    linkcolor=red,          
    citecolor=blue,        
    filecolor=magenta,      
    urlcolor=blue           
}
\usepackage[all]{hypcap} 
\usepackage{multirow}
\usepackage{multicol}

\newcolumntype{M}[1]{>{\centering\arraybackslash}p{#1}}

\usepackage{natbib}
\setlist[description]{leftmargin=\parindent,labelindent=\parindent}

\usepackage[normalem]{ulem} 

\newcommand{\nc}{\newcommand}

\newcommand{\muB}{\mu_{\text{B}}}
\newcommand{\nB}{n_{\text{B}}}
\newcommand{\muI}{\mu_{\text{I}}}

\newcommand{\muR}{\mu_{\text{R}}}

\newcommand{\mhat}{\hat{m}}
\newcommand{\uhat}{\hat{u}}
\newcommand{\vhat}{\hat{v}}

\nc{\beq}{\begin{equation}}
\nc{\eeq}{\end{equation}}
\nc{\beqa}{\begin{eqnarray}}  
\nc{\eeqa}{\end{eqnarray}}  
\nc{\bit}{\begin{itemize}}  
\nc{\eit}{\end{itemize}}

\newcommand{\ie}{{\it i.e.}}

\usepackage{floatrow}
\newfloatcommand{capbtabbox}{table}[][\FBwidth]

\usepackage{blindtext}

\allowdisplaybreaks

\title{ 
{\bf Approaching Stable Quark Matter}
\author{\large Yang Bai and Ting-Kuo Chen}
\date{\small \it 
Department of Physics, University of Wisconsin-Madison, Madison, WI 53706, USA
}
}

\begin{document}

\maketitle

\setlength{\parskip}{0.2ex}

\begin{abstract}
The determination of whether the ground state of baryon matter in Quantum Chromodynamics (QCD) is the ordinary nucleus or a quark matter state remains a long-standing question in physics. A critical parameter in this investigation is the bag parameter $B$, which quantifies the QCD vacuum energy and can be computed using nonperturbative methods such as Lattice QCD (LQCD). By combining the equation of state derived from perturbative QCD (pQCD) with the bag parameter to fit the LQCD-simulated data for isospin-dense matter, we address the stability of quark matter within the LQCD+pQCD framework. Our findings suggest that the current data imposes an upper bound on $B^{1/4} \lesssim 160$~MeV, approaching a conclusive statement on quark matter stability. Given the lower bound on $B$ from the quark condensate contribution to the vacuum energy, the stable 2-flavor quark matter remains possible, whereas the stable 2+1-flavor quark matter is excluded, assuming complete deconfinement and chiral-symmetry restoration and the reliability of pQCD at baryon chemical potentials around the proton mass. Additionally, we derive more general thermodynamic bounds on the quark matter energy-per-baryon and $B$, which, while weaker, provide complementary insights.
\end{abstract}

\thispagestyle{empty}  
\newpage    
\setcounter{page}{1}  

\begingroup
\hypersetup{linkcolor=black,linktocpage}
\tableofcontents
\endgroup

\newpage

\section{Introduction}\label{sec:intro}

It has long been thought that the global ground state of the baryon matter in quantum chromodynamics (QCD) should be the hadron state, while the quark matter state could only be accessible at environments of high temperature (such as the early universe or ion colliders) or high density (such as the core of a massive neutron star). However, the possible form of quark matter existence as a solitonic ``bag'' in the zero-pressure vacuum, also known as the ``quark nugget'', has long ago been hypothesized and studied extensively by physicists such as Bodmer~\cite{Bodmer:1971we}, Lee and collaborators~\cite{Lee:1974kn,Lee:1974uu,Friedberg:1976eg,Friedberg:1977xf}, and Witten~\cite{Witten:1984rs}, and still receives wide attention today. In contrast to the possible quark matter at the core of a neutron star, whose stability is supported by the external pressure, the stability of a quark nugget relies entirely on the energy configuration of the matter itself. Usually, a quark matter state is considered stable if its energy per baryon $\epsilon/\nB$ is less than around $930$~MeV, the averaged mass per nucleon of the iron element. So far whether this is true still remains an open question. The question of quark matter stability is of major importance since the implied stable quark nuggets could potentially extend the periodic table by introducing ``exotic nuclei" with very large atomic and atomic mass numbers. Furthermore, stable quark nuggets, if produced in the early universe, could serve as a compelling dark matter candidate~\cite{Witten:1984rs}. One such study of quark matter stability based on phenomenological models can be found in Ref.~\cite{Dexheimer:2013eua}. On the other hand, we aim to address this question using the fundamental QCD, which should be more robust than the known phenomenological models.

To investigate this question, one can start from the high-density limit, within which the thermodynamic equation of state (EOS) of the quark matter phase can be calculated based on perturbative QCD (pQCD)~\cite{Freedman:1976xs,Freedman:1976ub,Baluni:1977ms}. For instance, Ref.~\cite{Kurkela:2009gj} has performed such calculations up to $\mathcal{O}(\alpha_s^2)$ including the strange quark mass, which we mainly follow in this work. We note that partial results of the $\mathcal{O}(\alpha_s^3)$ calculations are available in Refs.~\cite{Gorda:2023mkk,Karkkainen:2025nkz}, which we will discuss later in this study. The pressure from pQCD calculations is a positive function of the baryon chemical potential $\muB$. The additional formation of color superconducting (CS) phase from Bardeen-Cooper-Schrieffer condensates provides also positive but negligible pressure to the quark matter system in the perturbative regime. To achieve a quark matter state with zero external pressure, an additional negative contribution to the total pressure is required. This contribution could arise from the vacuum energy difference between the confined (chiral-symmetry breaking) and deconfined (chiral-symmetry restored) QCD phases, which is nonperturbative in nature.

In the literature, this vacuum energy difference is often collectively described by the so-called ``bag parameter'' ($B$), which was introduced as a phenomenological parameter in the MIT bag model~\cite{Chodos:1974je,Chodos:1974pn}. More formally, this parameter can be defined as
\beqa\label{eq:trace_anomaly}
\hspace{3cm} B \equiv -\frac{1}{4}\,\langle \Theta^\mu_{\;\mu} \rangle ~, \qquad \mbox{with}\quad \Theta^\mu_{\;\mu} = \frac{\beta(g)}{2g}G_a^{\mu\nu}G^a_{\mu\nu} + \sum_q\,m_q\,[1+\gamma_m(g)]\,\overline{q}q ~.
\eeqa
Here, $\Theta^\mu_{\;\mu}$ is the trace anomaly that depends on the renormalized fields and the QCD gauge coupling $g$~\cite{Collins:1976yq,Nielsen:1977sy}; $\beta(g)$ is the beta-function of the gauge coupling and $\gamma_m(g)$ is the quark mass anomalous dimension. The vacuum expectation value of the trace anomaly operator is defined at zero temperature and density in the confined vacuum. One way to obtain $B$ is to infer the values of the condensates from other observables (see for example Refs.~\cite{Gell-Mann:1968hlm,Shifman:1978bx,Ioffe:2002be,Borsanyi:2012zv,McNeile:2012xh,Narison:2018dcr,Harnett:2021zug}); however, this approach suffers from large systematic uncertainties, which we will discuss in more detail in Section~\ref{sec:combine}.  An alternative approach is to directly perform nonperturbative calculations of the thermodynamic properties of the baryon matter system, and lattice QCD (LQCD) is by far the most unique tool for this task.

In order to explore the transition from the hadron phase to the quark matter phase, one can either probe along the temperature direction or along the chemical potential direction (see for example Ref.~\cite{Guenther:2020jwe} for a review on the QCD phase diagram). In this work, we focus on the latter and will comment on the former in Section~\ref{sec:conclusions}. It is well known that simulating a system at finite $\muB$ (LQCD$_{\rm B}$) suffers from the fermion sign problem~\cite{deForcrand:2009zkb}, making it challenging to extend the study to a high $\muB$. On the other hand, the isospin-dense matter system (LQCD$_{\rm I}$) is free of this problem (see Section~\ref{sec:lattice}), and thus can be studied up to a high isospin chemical potential $\muI$ to explore the deconfined quark matter phase. Recently, Refs.~\cite{Abbott:2023coj,Abbott:2024vhj} have performed LQCD simulations on a system of 6144 pions up to $\muI \sim 3.2$~GeV at zero temperature, in which they have identified the first-order phase transition to a Bose-Einstein condensate phase and a crossover to the CS phase~\cite{Son:2000xc,Son:2000by} through measuring the pressure $p$, energy density $\epsilon$, and speed of sound $c_s$ of the system. Based on the next-to-leading-order (NLO) calculations of the CS contribution to the EOS derived in Ref.~\cite{Fujimoto:2023mvc}, Ref.~\cite{Fujimoto:2024pcd} has performed a matching between the LQCD$_{\rm I}$ data and the pQCD+CS calculations to estimate the CS gap parameter. In this work, we take into account the $B$ parameter together with pQCD+CS to analyze the LQCD$_{\rm I}$ data. This will not only provide a more complete description of the system, but also potentially examine the stability of quark matter.

To investigate the quark matter stability problem using this LQCD+pQCD framework, we will take two steps. The first step is to identify the stability condition based on the pQCD+CS+$B$ calculations of the quark matter properties. In addition to $B$, another important factor of the calculations is the renormalization scale parameter $X$, defined as the ratio of the renormalization scale over a weighted sum of the quark chemical potentials (see Eq.~\eqref{eq:X-param}). The choice of $X$ has a great influence on the predictions in the low $\muB$ regime, which is the most relevant regime for quark matter stability. As a result, we treat the framework as a two-parameter model of $X$ and $B$ and explore the parameter space for stable quark matter. For the second step, we perform a two-parameter fit on the LQCD$_{\rm I}$ data with the pQCD+CS+$B$ framework to extract the information about the bag parameter sensitivity. By combining the results from the two steps, we can then determine whether the measured quark matter properties from LQCD$_{\rm I}$ indicate the existence of stable quark matter. Moreover, one can distinguish between the strange and strangeless quark matter scenarios, which is also a question of great interest in the studies of quark matter.

Nevertheless, as we will discuss later on, this LQCD+pQCD framework is only valid for $X > 1.42$, since perturbative calculations will break down for $\muB \leq 930$~MeV with $X \leq 1.42$. As a result, we further apply a parallel framework based on general thermodynamic constraints proposed in Ref.~\cite{Komoltsev:2021jzg} to obtain a less stringent but more robust bound that can be applied to the entire parameter space. In contrast to the LQCD+pQCD framework, which requires a full knowledge of the EOS, this thermodynamic framework requires only two reference points to produce the bounds. Though it certainly cannot provide as predictive constraints as the LQCD+pQCD framework, the thermodynamic framework offers a complementary approach to investigate the problem of quark matter stability in a more general and reliable way.

Our paper is organized as follows. In Section~\ref{sec:classification}, we classify two possible interrelations between the hadron and quark matter EOS's and discuss their implications for quark matter stability. In Section~\ref{sec:pQCD}, we derive the quark matter stability conditions based on pQCD calculations up to $\mathcal{O}(\alpha_s^2)$, with the relevant formulas and integral approximations summarized in Appendices~\ref{sec:quark}, \ref{sec:integral}, and \ref{sec:CFL-CS}. The LQCD$_{\rm I}$ data is analyzed in Section~\ref{sec:lattice} with the pQCD+CS+$B$ framework and used to infer the current and projected quark matter stability constraints on the $(X,B^{1/4})$ parameter space in Section~\ref{sec:combine}, with some of the details given in Appendix~\ref{sec:lattice_details}. In Section~\ref{sec:thermo}, we introduce the complementary approach to explore quark matter stability using general thermodynamic conditions, with the details given in Appendices~\ref{sec:bound} and \ref{sec:EOS}. Finally, we discuss and conclude our study in Section~\ref{sec:conclusions}.

\section{Phases of baryon matter}\label{sec:classification}
At zero temperature, a type of matter carrying a nonzero baryon number can be described by the grand potential per volume 
\beqa
- \frac{\Omega}{V} = p  = \muB \,\nB - \epsilon ~.
\label{eq:pressure-energy}
\eeqa
Here, $p$ and $\epsilon$ are the pressure and energy density of the system, while $\muB$ and $\nB$ are the baryon chemical potential and number density. The pressure $p(\muB)$ is a continuous function of the chemical potential and satisfies the basic thermodynamic relation $\partial p(\muB)/\partial\muB = \nB$ with $\nB \ge 0$. The stability of the thermodynamic system requires $\partial \nB/ \partial \muB \ge 0$~\cite{Callen:450289}, while the causality condition requires that the speed of sound $c_s$ be less than the speed of light or $c_s^{-2} \equiv (\muB/\nB)(\partial \nB/ \partial \muB) \ge 1$.

\begin{figure}[ht!]
    \centering
    \includegraphics[width=0.465\textwidth]{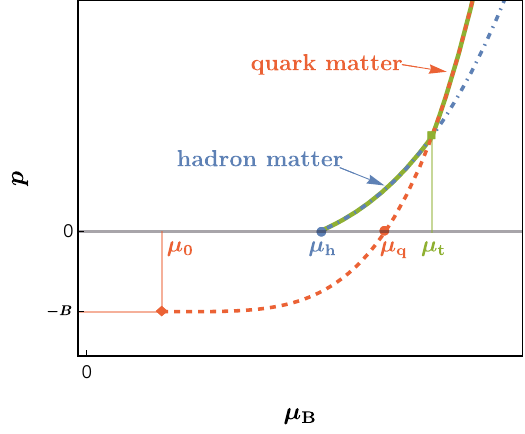} \hspace{8mm}
    \includegraphics[width=0.465\textwidth]{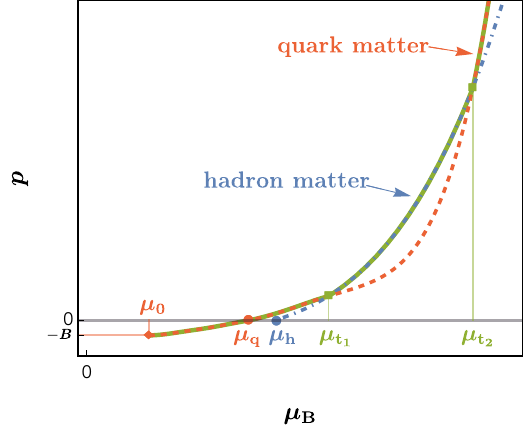}
    \caption{ \textit{Left panel:} A schematic plot of the $p(\muB)$ curves of the hadron matter phase and the quark matter phase at zero temperature, when there is only one intersection at $\muB=\mu_{\rm t}$ (green square point) between the two phases. The blue circle point at $\muB=\mu_{\rm h}$ with $p=0$ gives the energy per baryon of the (free) hadron matter as $\mu_{\rm h}$, while the red circle point at $\muB=\mu_{\rm q}$ gives that of the quark matter. The lower endpoint of the quark matter phase at $\muB=\mu_0$ and $p = - B$ has $\nB = 0$. The globally stable baryon matter phase (green solid line) follows the trajectory of the maximum $p$ at any given $\muB$. \textit{Right panel:} The same as the left panel but with two intersections at $\muB=\mu_{{\rm t}_1}$ and $\mu_{{\rm t}_2}$ between the two phases.
    }
    \label{fig:construction}
\end{figure}

In this paper, we will mainly study two possible phases of the baryon matter: the hadron matter phase and the quark matter phase (including the possible color superconductivity (CS) phase). Treating $\muB$ as the only intrinsic thermodynamic variable, we show a schematic plot in Figure~\ref{fig:construction} of the pressure (negative of the grand potential per volume) as a function of the baryon chemical potential. Due to the thermodynamic stability constraint~\cite{Callen:450289}, the pressure must be a convex function of $\muB$. For a fixed value of $\muB$, the phase with a larger $p$ is more stable. 

The ordinary hadron matter phase (blue dot-dashed line) starts from a nonzero $\muB=\mu_{\rm h}$ (bounded from below by the baryon mass), where both the pressure and the number density are zero. As for the quark matter phase (red dashed line), the starting chemical potential $\mu_0$\footnote{As we will discuss more in Section~\ref{sec:pQCD} and Appendix~\ref{sec:quark}, $\mu_0$ results mathematically from the breakdown of pQCD calculations when $\alpha_s$ becomes too large. While there is no rigorous proof, $\mu_0$ could also play a similar physical role for the quark matter as $\mu_{\rm h}$ for the hadron matter, the latter of which determines the minimum energy per baryon needed to produce a hadron state.} could lie at a lower value, where $\nB$ is zero but the pressure is non-positive, $p = - B$ with $B \ge 0$. As $\muB$ increases, there will be a zero-pressure point at $\muB=\mu_{\rm q}$ where the free quark matter state can exist with its energy per baryon given by $\epsilon/\nB = \mu_{\rm q}$ as derived from Eq.~\eqref{eq:pressure-energy}.

The $p(\muB)$ curves of the two phases are expected to have at least one intersection. In the high $\muB$ limit, one anticipates the quark matter phase (including the CS phase) to be the ultimate phase, and the hadron matter phase starting from $\muB=\mu_{\rm h}$ will eventually flow to the quark matter phase. In general, the $p(\muB)$ curves of the two phases can intersect multiple times. For simplicity, we will consider the situation with only one intersection at $\mu_{\rm t}$ (shown in the left panel of Figure~\ref{fig:construction}) and that with two at $\mu_{{\rm t}_{1,2}}$ (shown in the right panel of Figure~\ref{fig:construction}). Using the Maxwell construction to patch the curves of the two different phase,\footnote{Alternatively, the Gibbs construction further considers the electric charge chemical potential of the thermodynamic system.} both the chemical potentials and pressures of the two phases should be continuous at the transition boundaries, \ie, $\mu_1=\mu_2=\mu_{\rm t}$ and $p_1(\mu_{\rm t})=p_2(\mu_{\rm t})$.
For the case with one intersection in the left panel, one inevitably has $\mu_{\rm q} > \mu_{\rm h}$ or $\epsilon/\nB|_{\rm q} > \epsilon/\nB|_{\rm h}$, indicating an unstable quark matter state. For a fixed value of $\mu_{\rm t}$, $\mu_{\rm q}$ will increase along with increasing $B$, making the quark matter state more unstable. Conversely, $\mu_{\rm q} < \mu_{\rm h}$ for the case of two intersections in the right panel, resulting in a more stable quark matter state compared to the ordinary hadron matter state.

Here, the information of the different phases is derived from the gross thermodynamic properties of the system, which ignores the details of the microscopic physics. Even though the phase with a larger value of $p$ for a given $\muB$ is the globally stable phase, the phase with a smaller value of $p$ could still exist with a long lifetime, potentially longer than the age of the Universe. Using the right panel of Figure~\ref{fig:construction} as an example, the  hadron matter phase interval stretching from $\muB=\mu_{\rm h}$ to $\mu_{{\rm t}_1}$ is less stable than the quark matter phase. However, the actual nuclear matter in the hadron matter phase does not exist as a homogeneous and isotropic fluid but as elements with well-defined atomic structures. Within these finite-size structures, the ground state quark matter (also known as the ``quark nugget'') becomes less stable due to the finite-size effect and its lifetime should in general depend on the atomic mass number $A$. As long as the minimum $A$ at which the quark matter ground state is the true QCD ground state is greater than the maximum known $A$ (approximately 300) for the ordinary nuclear matter, \ie, $A_{\rm min} \gtrsim 300$, the decay of the ordinary nuclear matter can be avoided while stable quark nuggets can still exist for $A > A_{\rm min}$. Similarly, the quark matter in the phase interval stretching from $\muB=\mu_{\rm q}$ to $\mu_{\rm t}$ in the left panel could have a long enough lifetime if it actually exists in the form of some finer structure.

Based on these phase diagrams, one could classify the existence scenarios of the neutron, quark, and hybrid stars. For the situation depicted in the left panel of Figure~\ref{fig:construction}, the core baryon chemical potential of the ordinary neutron stars made of only hadronic matter is bounded by the transition point $\muB=\mu_{\rm t}$, while hybrid stars with a quark matter core could form if the core baryon chemical potential exceeds $\mu_{\rm t}$. As for the case shown in the right panel, besides ordinary neutron and hybrid stars, quark stars could also form along the quark matter phase curve from $\muB=\mu_{\rm q}$ to $\mu_{{\rm t}_1}$. 

Before concluding this section, we comment on the possible scenario that the hadron and quark matter phases are connected through a smooth crossover. In this case, one should expect $p(\muB)=S(\muB)p_1(\muB) + [1-S(\muB)]p_2(\muB)$ with $S(\muB)\in[0,1]$ as an infinitely differentiable switching function between the two phases~\cite{Albright:2014gva,Kapusta:2021ney}. Since there is no clear boundary between the two phases, the simple picture of the quark matter pressure balancing the vacuum pressure only applies at asymptotically high $\muB$. In this study, we will assume that this picture is valid either due to the first-order nature of the phase transition or that the $\muB$ range of our interest is high enough to justify this assumption for a crossover phase transition.

\section{Quark matter stability from pQCD calculations}\label{sec:pQCD} 

At zero temperature and a high enough chemical potential, the grand potential or the pressure of the quark matter can be calculated perturbatively by treating the QCD coupling $\alpha_s$ as an expansion parameter~\cite{Freedman:1976xs,Freedman:1976ub,Baluni:1977ms,Kurkela:2009gj} and including the single quark loops, two-gluon irreducible vacuum diagrams, and the plasmon ring sum (see Ref.~\cite{Kurkela:2009gj} for details). In this section, we follow Ref.~\cite{Kurkela:2009gj} to calculate the thermodynamic properties of the quark matter system up to $\mathcal{O}(\alpha_s^2)$.\footnote{The plasmon ring sum contains terms of $\mathcal{O}(\alpha_s^2\ln\alpha_s)$ resulting from the resummations, though for simplicity we still refer to the nominal order as $\mathcal{O}(\alpha_s^2)$.} Together with the unknown bag parameter $B$, which can only be calculated nonperturbatively, we derive the stability conditions for the quark matter. 

We consider two different scenarios: the 2+1-flavor including a finite strange-quark mass and the massless 2-flavor systems. The relevant formulas are summarized in Appendix~\ref{sec:quark} for the convenience of the readers, with the approximations of the integral functions listed in Appendix~\ref{sec:integral}. Under the $\beta$-equilibrium (the effects of relaxing this condition will be discussed in Appendix~\ref{sec:quark}) and charge neutrality conditions, there is only one independent chemical potential of the system, which we choose to be $\mu_d \equiv \mu$ for both the 2+1- and 2-flavor scenarios. The baryon chemical potential is $\muB \equiv N_c(\sum_in_i\mu_i)/(\sum_in_i)$ with $N_c=3$ and summed over quark flavors, which in the massless limit are $\muB = \mu_u + \mu_d + \mu_s$ and $\muB = \mu_u + 2\mu_d$, respectively for the two scenarios. Moreover, as we will discuss more in Section~\ref{sec:lattice}, we also consider the $u\overline{d}$ isospin-dense matter system, in which $\mu_u = -\mu_d = \mu$ with the isospin chemical potential $\mu_{\rm I} \equiv \mu_u - \mu_d = 2 \mu$.

The pressure calculated from pQCD, $p_{\rm pQCD}\big(\muB, \alpha_s(\muR),  m_s(\muR) \big)$, is a function of the baryon chemical potential, QCD gauge coupling, and renormalized strange quark mass. Here $\muR$ is the renormalization scale. When $\muR\lesssim800$\,MeV, the QCD interaction becomes strong and causes increased uncertainties in the pQCD calculations from ignoring higher-order contributions in $\alpha_s(\muR)$. On the other hand, one needs a proxy to relate the renormalization scale to the physical scale of the system (see Refs.~\cite{Schneider:2003uz,Kurkela:2009gj} for example). Given the uncertainty in the choice of $\muR$, we define a parameter $X$ to be the ratio of $\muR$ over a system-dependent weighted sum of the quark chemical potentials and will promote $X$ to a ``model parameter" when we consider the pQCD predictions. More specifically, we define $X$ for the different scenarios as

\begin{equation}\label{eq:X-param}
    X = \begin{cases}
        \dfrac{\muR}{\frac{2}{N_c}\left(\mu_u+\mu_d+\mu_s\right)} &~ \text{[2+1-flavor]} \\[3ex]
        \dfrac{\muR}{\frac{2}{N_c}\left(\mu_u+2\mu_d\right)} &~ \text{[2-flavor]} \\[3ex]
        \dfrac{\muR}{\mu_u-\mu_d} &~ \text{[isospin-dense]}
    \end{cases} ~,
\end{equation}
with $N_c = 3$. This is roughly the ratio of $\muR$ over twice of the averaged quark chemical potential.  Based on this definition, the pressure from pQCD is $p_{\rm pQCD}(\muB, X)$, a function of two variables. 

In addition to the perturbative contributions, it is long known that color superconductivity (CS) in the color antitriplet $qq$ channel could also arise in the finite-density quark matter, which effectively contributes an extra positive term to the quark matter pressure. Moreover, a similar color singlet $\overline{q}q$ phase specific to the isospin-dense matter could also arise at finite density. Refs.~\cite{Fujimoto:2023mvc,Fujimoto:2024pcd} have calculated the associated pressures of these condensates $p_{\rm CS}$ up to the NLO, from which we quote the formula and present it in Eq.~\eqref{eq:P_SC}. Because of the different color factors, the color antitriplet channel is much more suppressed compared to the pQCD contributions, while the color singlet channel plays a major role in the analysis of the isospin-dense matter (see Section~\ref{sec:lattice}). We have found that including $p_{\rm CS}$ for the color antitriplet channel has a negligible effect on our final results.

\begin{figure}[th!]
    \centering
    \includegraphics[width=0.47\linewidth]{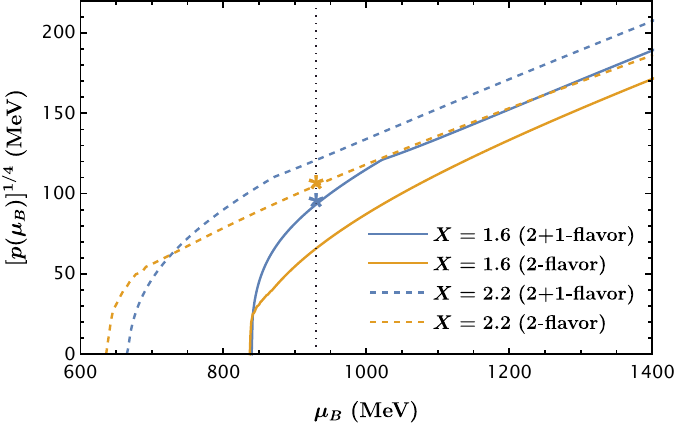} \hspace{0.5cm}
    \includegraphics[width=0.47\linewidth]{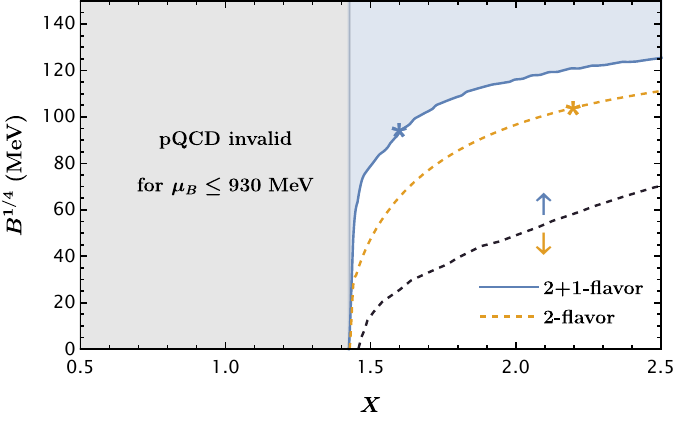}
    \caption{{\it Left panel:} The $p(\muB, X) = p_{\rm pQCD} + p_{\rm CS}$ distributions of the benchmarks with $X=1.6$ and $2.2$ for the 2+1- and 2-flavor systems. The black dotted vertical line stands for $\muB = 930$~MeV, whose intersections with the $p$ distributions imply the upper bounds on the bag parameter $B$ that allow stable quark matter. {\it Right panel:} The parameter region (white) of stable quark matter with trustable pQCD calculations. The regions above the blue solid and orange dashed lines exclude the existence of stable 2+1- and 2-flavor quark matter, respectively. The pQCD calculation breaks down for $\muB \leq 930$~MeV in the shaded gray region of $X \leq 1.42$ (see Appendix~\ref{sec:quark} for details) and thus cannot give conclusive remarks about the stability of the quark matter. Above (below) the black dashed line, the 2+1-flavor (2-flavor) quark matter is more energetically stable, as indicated by the blue (orange) arrows. The two asterisks correspond to the two asterisks marked in the left panel.
    \label{fig:pQCD_constraint}
    }
\end{figure}

For a quark matter system to be stable, the matter pressure (comprising both the pQCD and CS contributions) must balance the bag pressure $-B$ at a baryon chemical potential $\muB \leq 930$~MeV,\footnote{One could relax this condition by using the free nucleon mass $m_p \approx 938$~MeV instead, assuming a very long tunneling time to form a nucleus. This brings a negligible change to the range of the relevant parameter space.} the energy per baryon of the iron element. We illustrate this with two benchmarks of $X=1.6$ and $2.2$ for both the 2+1- and 2-flavor systems by plotting their $p(\muB, X) = p_{\rm pQCD} + p_{\rm CS}$ distributions in the left panel of Figure~\ref{fig:pQCD_constraint}. The black dotted vertical line in the plot stands for $\muB = 930$~MeV, and its intersections with the $p$ distributions imply the upper bounds on the bag parameter $B$ that allow stable quark matter for the different systems. As discussed in Ref.~\cite{Kurkela:2009gj}, the region with $p\simeq0$ is also where $\alpha_s$ becomes large such that the pQCD results become unreliable, which is manifested in that $\nB(\muB<\mu_0)<0$ and $p(\muB=\mu_0)=0$ (see Appendix~\ref{sec:quark} for details). Depending on the choice of the $X$ parameter, either the 2+1- or 2-flavor quark matter could be more stable than the other one. To guide our discussion, we use the asterisks to mark the locations of $p^{1/4}$ at $\muB = 930$~MeV for the $X=1.6$ benchmark of the 2+1-flavor scenario and the $X=2.2$ benchmark of the 2-flavor scenario.     

In the right panel of Figure~\ref{fig:pQCD_constraint}, we show the consequent constraint plot on the $X$-$B^{1/4}$ plane that implies stable quark matter. The gray shaded region with $X \leq 1.42$ is the region that the pQCD calculations break down for $\muB \leq 930$~MeV since $\mu_0\leq930$~MeV. In this region, no conclusive remarks based on pQCD can be made about the stability of the quark matter. For the region with $X > 1.42$, the blue and orange lines denote the boundary of stable 2+1- and 2-flavor quark matter, respectively. For the regions between the blue solid and orange dashed lines, only the 2+1-flavor quark matter can be stable. The black dashed line shows the boundary of the relative stability between the 2+1-flavor and 2-flavor quark matter: the region above (below) the black dashed line indicates that the 2+1-flavor (2-flavor) quark matter is more energetically stable. Also note that there is some mild numerical uncertainty in the region around $X = 1.42$, which does not change the general boundary feature. From this plot, one can see that for most $X$ values (except the uncertain region around $X = 1.42$), the 2-flavor scenario is more stable when $B^{1/4}$ is small (below the black dashed line) and the other way around when $B^{1/4}$ is large (above the black dashed line). This is due to the fact that the strange flavor only becomes more populated when $\muB$ is high, thus contributing more to the pressure of the 2+1-flavor system, but will be suppressed in the low-$\muB$ regime where the strange quark mass matters more.

Finally, we make a comment on the perturbative convergence of the calculations and point out a possible caveat of the stability analysis presented in this section, with the details given in Appendix~\ref{sec:quark}. In the $\muB$ range of our interest, we find that $\alpha_s(\muB)\lesssim0.6$ for $X\geq1.47$ for both the 2- and 2+1-flavor scenarios. The fact that $(\frac{\alpha}{\pi}) < 1$ should indicate the applicability of pQCD subject to a complication of different $SU(3)_C$ group factors showing up at different loop orders. As we show in the Appendix using a simple massless three-flavor scenario with $X=2$, the $\mathcal{O}[(\frac{\alpha}{\pi})^1]$ and $\mathcal{O}[(\frac{\alpha}{\pi})^2]$ corrections, though being comparable in magnitude, are indeed subdominant compared to the $\mathcal{O}[(\frac{\alpha}{\pi})^0]$ contribution and of the same negative sign. Specifically, $\alpha_s(\muB=930~{\rm MeV})=0.42$, clearly satisfying the condition $(\frac{\alpha}{\pi})<1$. We further show in the Appendix that if the $\mathcal{O}[(\frac{\alpha}{\pi})^3]$ correction turns out to completely cancel the $\mathcal{O}[(\frac{\alpha}{\pi})^2]$ correction, \ie, leaving only the $\mathcal{O}[(\frac{\alpha}{\pi})^1]$ correction intact, our conclusion on quark matter stability still remains robust (see Section~\ref{sec:combine} for the conclusion based on the $[(\frac{\alpha}{\pi})^2]$ calculations). Based on the partial results presented in Ref.~\cite{Gorda:2023mkk}, such a dramatic cancellation seems very unlikely to happen; moreover, the increasing trend of $\mu_0$ for high $X$ values suggests that higher-order corrections will further solidify our conclusion in this paper. However, until the complete calculations are done~\cite{Karkkainen:2025nkz}, there is no guarantee that the $\mathcal{O}[(\frac{\alpha}{\pi})^3]$ correction will not be larger than the sum of the $\mathcal{O}[(\frac{\alpha}{\pi})^1]$ and $\mathcal{O}[(\frac{\alpha}{\pi})^2]$ corrections with an opposite sign and invalidate the predictions made in this study.

\section{Bag parameter from Lattice QCD calculation of isospin-dense matter}\label{sec:lattice} 

To determine whether quark matter is stable, an important parameter to consider is the bag parameter $B$, the calculation of which requires a nonperturbative tool such as Lattice QCD (LQCD). In this section, we highlight that recent LQCD simulation results can be used to infer the value of $B$.
  
One may attempt to directly use LQCD to simulate the baryon-dense system. However, the notorious sign problem at nonzero $\muB$~\cite{deForcrand:2009zkb} prohibits direct application. On the other hand, the isospin-dense matter system in the isospin-symmetric limit is free of this concern, as utilized in Ref.~\cite{Cohen:2003ut} to impose a QCD inequality bound on the baryon matter. To see this, one can write down the partition function of the system with $\mu_d = - \mu_u = \mu$ for the isospin-dense system in the Euclidean spacetime: 
\begin{equation}
\begin{aligned}
    Z_{\rm I}(\mu) &= \int\mathcal{D}[A]\det\mathcal{D}(-\mu)\det\mathcal{D}(\mu)e^{-S_G} = \int\mathcal{D}[A]\,\left\vert \det\mathcal{D}(\mu)\right\vert^2e^{-S_G} ~,
\end{aligned}
\end{equation}
where $A$ is the gluon field, $S_G$ is the gluon field action, and $\mathcal{D}(\mu)=\slashed{D}+m_q-\mu\gamma_0$ is the Dirac operator. It is clear that $Z_{\rm I}$ is free of the sign problem as it only depends on the norm $\vert\det\mathcal{D}\vert$, thus allowing one to perform LQCD simulations for a wide range of $\mu$ values. Consequently, one can derive the following bound on the baryon matter pressure $p_{\rm B}$ using the aforementioned QCD inequality based on the isospin-dense matter pressure $p_{\rm I}$,
\begin{equation}
    p_{\rm B}(\mu_{\rm B}) \leq p_{\rm I}\left(\mu_{\rm I}=\frac{2\mu_{\rm B}}{N_c}\right) ~.
\end{equation}
This has been investigated in Ref.~\cite{Fujimoto:2023unl} using the LQCD dataset from Ref.~\cite{Abbott:2023coj} (we will discuss it more later). Following their methodology but using the more updated LQCD data from Ref.~\cite{Abbott:2024vhj}, we find that the resulting bound [$p_{\rm B}(\mu_{\rm B}=930~{\rm MeV})\leq(210~{\rm MeV})^4$], though robust, is much looser than the pressure values derived using pQCD for the $X$ values we consider [$p_{\rm B}(\mu_{\rm B}=930~{\rm MeV})\leq(150~{\rm MeV})^4$], and thus we will not discuss them further for our purpose of studying quark matter stability. Nevertheless, this bound could be useful for the investigation of quark matter scenarios beyond the capability of pQCD.

Recently, the NPLQCD collaboration performed a simulation on a large isospin-dense system up to $\mu \sim 1.6$~GeV at $T\sim\mathcal{O}(10)~{\rm MeV}$ (which should be low enough to be considered as ``cold'') and made their data publicly available in Ref.~\cite{Abbott:2024vhj} (their earlier simulations can be found in Ref.~\cite{Abbott:2023coj}). Notably, the isospin-dense matter system shares the same bag parameter $B$ as the baryon-dense matter system. We will analyze the LQCD data for the isospin-dense system to obtain the value of or the bound on $B$.  

\begin{figure}[bht!]
    \centering
    \includegraphics[width=0.32\linewidth]{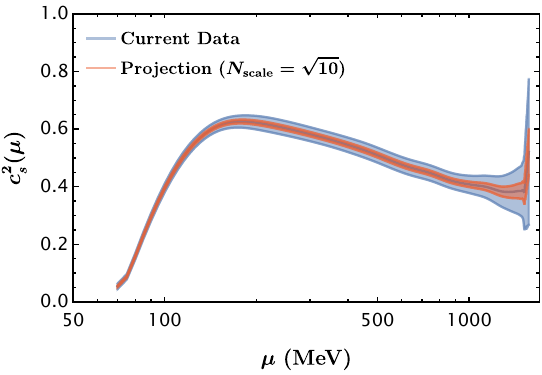} \hspace{1.5mm}
    \includegraphics[width=0.32\linewidth]{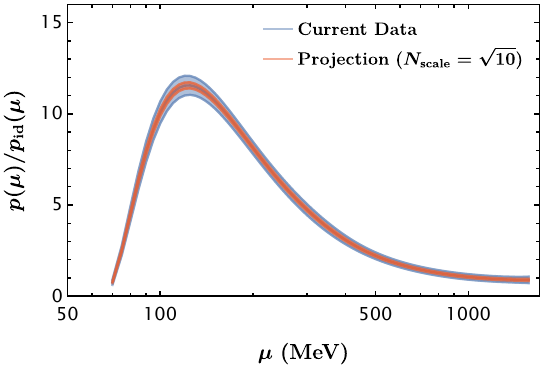}\hspace{1.5mm}
    \includegraphics[width=0.32\linewidth]{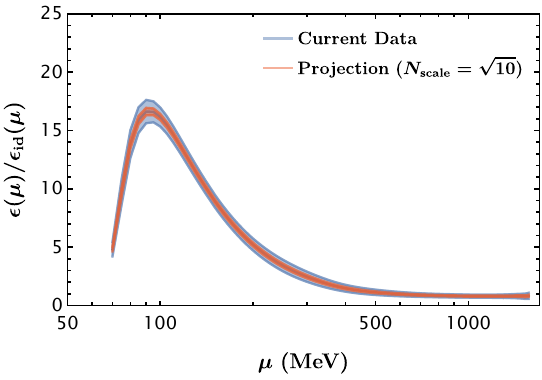}
    \caption{Reproduced distributions for $c_s^2(\mu)$, $p(\mu)/p_{\rm id}(\mu)$, and $\epsilon(\mu)/\epsilon_{\rm id}(\mu)$ using the publicly available LQCD data in Ref.~\cite{Abbott:2024vhj}. We also show the projected distributions with an expected decrease of $1/N_{\rm scaled}=1/\sqrt{10}$ in the bin-wise uncertainties.
    }
    \label{fig:lattice}
\end{figure}

To start with, we use the LQCD data provided in Ref.~\cite{Abbott:2024vhj} to reproduce the distributions of three quantities (shown in Figure~\ref{fig:lattice}): the square of the speed of sound $c_s^2(\mu)$, normalized pressure $p(\mu)/p_{\rm id}(\mu)$, and normalized energy density $\epsilon(\mu)/\epsilon_{\rm id}(\mu)$, where $p_{\rm id}(\mu) = \epsilon_{\rm id}(\mu)/3=3N_f\mu^4/(12\pi^2)$ for the ideal Fermi gas and $N_f = 2$. For our later purpose to obtain the projected bound on $B$, we also show the projected distributions with an expected decrease of $1/N_{\rm scaled}=1/\sqrt{10}$ in the bin-wise uncertainties assuming that 10 times more ensembles can be made in the future simulations.

As first pointed out in Ref.~\cite{Fujimoto:2024pcd}, comparing the LQCD data to the pQCD calculations at finite density will shed much light on our understanding of QCD. While Ref.~\cite{Fujimoto:2024pcd} focuses on the role played by the CS condensates, we are interested in the inference of the bag parameter through comparing the LQCD and pQCD calculations, which would in turn imply the stability of the quark matter. Such is possible since $B$ is universal to both the baryon and isospin-dense matter, an important point that allows one to study the latter free of the sign problem with LQCD to infer about the former, as we have mentioned earlier. Given the crossover nature of the phase transition in the isospin-dense system, one might question the validity of introducing $B$ and modeling the system using pQCD; however, as we will discuss later, since we only analyze the data with $\mu> 1000$~GeV (an alternative analysis of data with $\mu> 800$~MeV is given in Appendix~\ref{sec:lattice_details}), perturbativity should be valid while the vacuum description with $B$ is asymptotically correct.
The pQCD model to fit the isospin-dense LQCD data is given by
\begin{equation}
    p(\mu,X,B) =  p_{\rm pQCD}(\mu,X) \,+\, p_{\rm CS}(\mu,X) \,-\, B ~,
\end{equation}
along with
\begin{equation}
    \epsilon = -p + \mu\frac{\partial p}{\partial \mu} ,~\qquad  c_s^2 = \frac{\partial p/\partial\mu}{\partial\epsilon/\partial\mu} ~,
\end{equation}
where the details of $p_{\rm pQCD}$ and $p_{\rm CS}$ can be found in Appendix~\ref{sec:quark} and Appendix~\ref{sec:CFL-CS}, respectively.\footnote{We note that in this work $p_{\rm QCD}$ is calculated up to the three-loop level, while $p_{\rm CS}$ is calculated up to the two-loop level diagrammatically. One can make the predictions more consistent by improving the calculations of $p_{\rm CS}$ to a higher-loop order in the future.} There are two independent model parameters to be fit: $X$ and $B$.

\begin{figure}[bht!]
    \centering
    \includegraphics[width=0.5\linewidth]{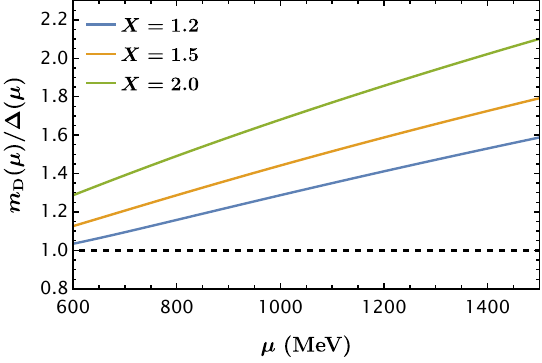}
    \caption{The ratio of the gluon Debye mass $m_{\rm D}(\mu)$ over the color-superconducting gap $\Delta(\mu)$ as a function of the quark chemical potential $\mu$ for $X = 1.2, 1.5, 2.0$. 
    }
    \label{fig:gap}
\end{figure}

In practice, one should trust the pQCD calculations only down to a certain $\mu = \mu_{\rm start}$. While loop-perturbativity in terms of the QCD gauge coupling is one of the main concerns, a more stringent bound is the validity of the gap formulas (see Eqs.~\eqref{eq:P_SC} and \eqref{eq:gap} in Appendix~\ref{sec:CFL-CS}), which requires that $\Delta < m_{\rm D} < \mu$, where $m_{\rm D} = (2\alpha_s \mu^2/\pi)^{1/2}$ is the gluon Debye mass~\cite{Pisarski:1999tv,Fujimoto:2024pcd} (see Ref.~\cite{Evans:1999at} for the effects of relaxing this condition). We show the ratio $m_{\rm D}(\mu)/\Delta(\mu)$ for a few $X$ values in Figure~\ref{fig:gap}. Since for the baryon matter, pQCD calculation is invalid for $\muB \leq 930$~MeV when $X \leq 1.42$, the hierarchy is justified for the $X$ range of our interest down to $\mu \gtrsim 600$~MeV. We make a conservative choice of $\mu_{\rm start} = 1000$~MeV for the main study, and present the results for $\mu_{\rm start} = 800$~MeV in Appendix~\ref{sec:lattice_details}. For $\mu < \mu_{\rm start}$, we simply assume zero discrepancies between the LQCD and pQCD calculations, which is a conservative treatment for constraining the bag parameter although including them may change the fitted central value of $B$. 

Superficially, the overall degrees of freedom of the data seem to be given by $300$($\mu$ bins)$\times3$ (observables) $= 900$. However, as pointed out in Ref.~\cite{Abbott:2024vhj}, the correlation among the neighboring $\mu$ bins and among the different observables greatly reduces the effective degrees of freedom of the data. This is reflected in the large hierarchy among the eigenvalues of the covariance matrix $\Sigma$ (which we show along with the associated eigen-observables in Figure~\ref{fig:Sigma_eigen} of Appendix~\ref{sec:lattice_details}), and thus we should truncate the fit at a certain number of the effective degrees of freedom $k$ to avoid overfitting. We note that the total $\chi^2$ will be dominated by the last few degrees of freedom, which carry the smallest eigenvalues of $\Sigma$. These degrees of freedom, assuming we are not overfitting the data, will provide the most stringent constraints on the parameter space.

\begin{figure}[hbt!]
    \centering
    \includegraphics[width=0.32\linewidth]{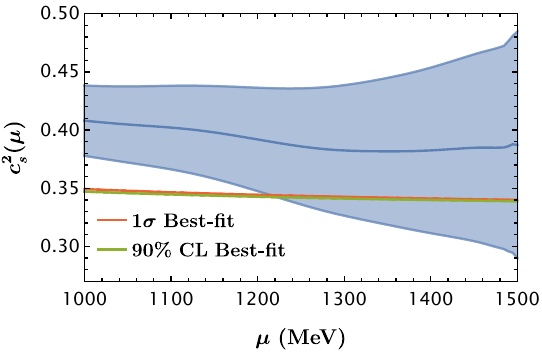} \hspace{1.5mm}
    \includegraphics[width=0.32\linewidth]{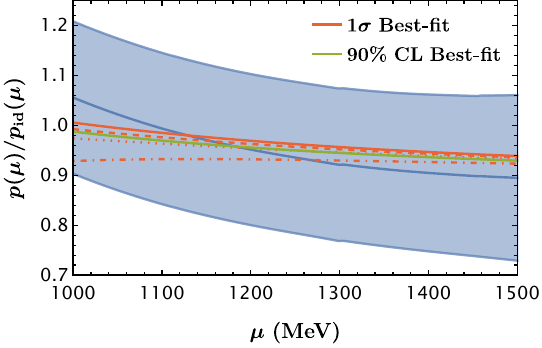}\hspace{1.5mm}
    \includegraphics[width=0.32\linewidth]{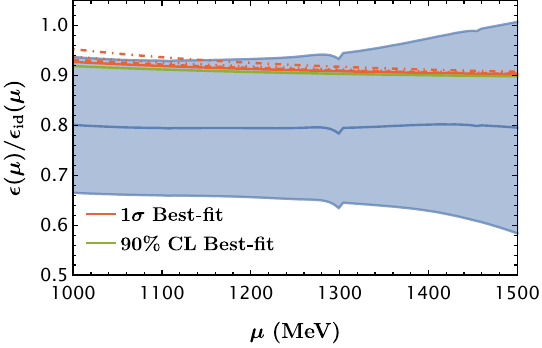}
    \caption{The $c_s^2$, $p/p_{\rm id}$, and $\epsilon/\epsilon_{\rm id}$ distributions of the best-fit points (red and green solid lines) with $\mu_{\rm start} = 1000$~MeV and $k = 49, 53$, whose benchmark parameters are $(X, B^{1/4}) = (1.66, 0~{\rm MeV})$ and $(1.78, 0~{\rm MeV})$. Also shown are the distributions of the benchmarks $(X, B^{1/4}) = (1.66, 161~{\rm MeV})$ (red dashed line), which sits on the $90\%$ CL contour around the $1\,\sigma$ ($k=49$) best-fit point, $(X, B^{1/4}) = (1.66, 200~{\rm MeV})$ (red dotted line), and $(1.66, 250~{\rm MeV})$ (red dot-dashed line).
    }
    \label{fig:fit_1000}
\end{figure}

For each $\mu_{\rm start}$, we determine the values of $k$ by requiring the best-fit point to sit just within the $1\sigma$ and $90\%$ confidence level (CL) contours for $k-2$ degrees of freedom. For $\mu_{\rm start} = 1000$~MeV, one has $k = 49$ and $53$ for the $1\sigma$ and $90\%$ CL best-fit conditions, respectively, whose benchmark parameters are $(X, B^{1/4}) = (1.66, 0~{\rm MeV})$ and $(X, B^{1/4}) = (1.78, 0~{\rm MeV})$. The LQCD data prefers a zero bag parameter based on the prior $B \ge 0$. This is mainly driven by the smaller energy density $\epsilon(\mu)$ of the LQCD data compared to the pQCD predictions (see the right panel of Figure~\ref{fig:fit_1000}). If we require a smaller confidence level that hints at overfitting the data, one has a smaller number of effective degrees of freedom $k$.

After determining the best-fit points and the corresponding values of $k$, we trace out the $90\%$ CL contour on the $(X, B^{1/4})$ plane, which will provide the bounds on both $X$ and $B$. In Figure~\ref{fig:fit_1000}, we show the $c_s^2$, $p/p_{\rm id}$, and $\epsilon/\epsilon_{\rm id}$ distributions of the two best-fit points (red and green solid lines) as well as those of the benchmark $(X, B^{1/4}) = (1.66, 161~{\rm MeV})$ (red dashed line), which sits on the $90\%$ CL contour around the $k = 49$ best-fit point. For illustration, we further show the predictions of two other benchmarks: $(X, B^{1/4}) = (1.66, 200~{\rm MeV})$ (red dotted line) and $(X, B^{1/4}) = (1.66, 250~{\rm MeV})$ (red dot-dashed line), with the latter clearly showing a discrepancy from the LQCD data. Note that $c_s^2$ does not rely on $B$ and thus the red lines completely overlap with one another in the left panel.

\section{Inferred quark matter stability} \label{sec:combine}

\begin{figure}[htb!]
    \centering
    \includegraphics[width=0.49\linewidth]{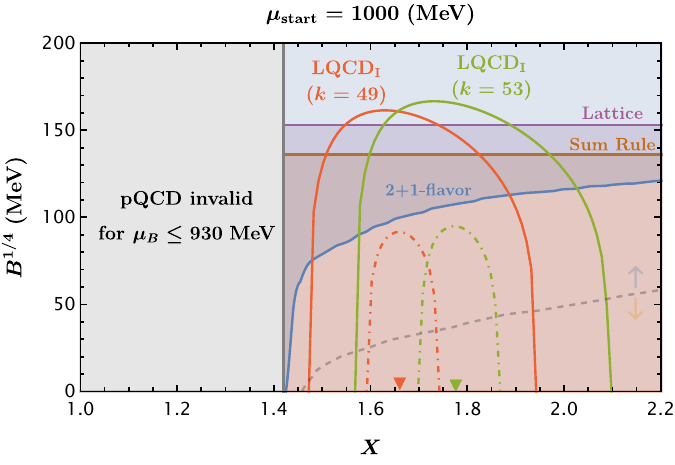}
    \includegraphics[width=0.49\linewidth]{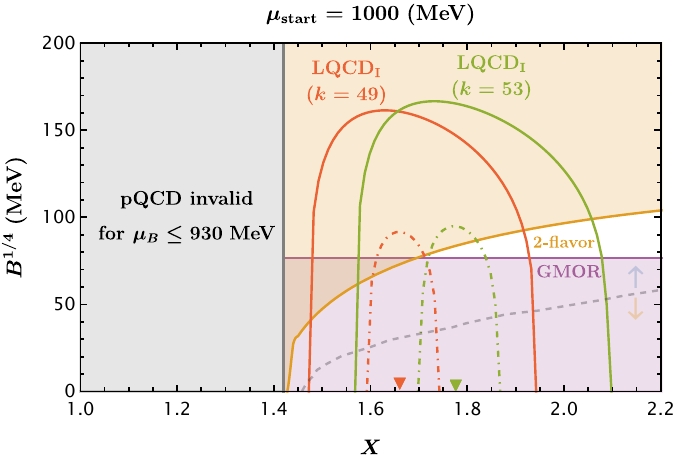}
    \caption{{\it Left panel:} A superposition of the allowed $X$-$B^{1/4}$ parameter space for the 2+1-flavor stable quark matter from the right panel of Figure~\ref{fig:pQCD_constraint} based on pQCD calculations and the constraints from $\mbox{LQCD}_{\rm I}$ simulation data. The red and green triangle points are the best-fit points for $k=49$ (the total $\chi^2$ at 1\,$\sigma$) and $k=53$ (the total $\chi^2$ at 90\% CL). The red and green solid lines are the 90\% CL contours around the best-fit points. The red and green dot-dashed lines are the ``projected limits" with the same values of $k=49$ (red) and $k=53$ (green) with a decrease in the LQCD uncertainties by a factor of $N_{\rm scale}=\sqrt{10}$. $\mu_{\rm start}=1000$~MeV is used for the LQCD data analysis. The lower bounds from Eq.~\eqref{eq:low_3-flavor} are also shown in the plot. {\it Right panel:} Same as the left panel but for the 2-flavor stable quark matter, with the lower bound from Eq.~\eqref{eq:low_2-flavor} also shown in the plot.
    }
    \label{fig:result_1000}
\end{figure}

In this section, we combine the constraints on $X$ and $B$ for stable quark matter based on pQCD calculations in Section~\ref{sec:pQCD} and those from analyzing the LQCD data with $\mu_{\rm start}=1000$~MeV in Section~\ref{sec:lattice}. The superposed result is summarized in Figure~\ref{fig:result_1000}. One can see that the best-fit points (red and green triangle points) to the LQCD data prefer a larger value of the renormalization scale parameter $X \approx 1.6-1.8$. The 90\% CL contours (red and green solid curves for $k=49$ and $k=53$, respectively) cover $X$ values from 1.5 to 2.1, which are within the trustable region of the pQCD calculations. This means that one could infer the stability of quark matter based on the $p(\muB)$ derived from pQCD and the bag parameter inferred from $\mbox{LQCD}_{\rm I}$, the LQCD calculations for the isospin-dense matter.   

Along the bag parameter $B$ direction, the current LQCD data prefers a zero value of $B$, which can be traced back to the $\epsilon(\mu)$ distribution shown in the right panel of Figure~\ref{fig:fit_1000} that is lower than the naively anticipated value of $B^{1/4} = \mathcal{O}(100\,\mbox{MeV})$ or the QCD scale. Interestingly, the 90\% CL constraints on the bag parameter from the $\mbox{LQCD}_{\rm I}$ data exert the limit $B^{1/4} \lesssim 160$~MeV, which is close to but above the threshold (80-120\,\mbox{MeV} for $X \in [1.5,2.1]$) for stable quark matter. Therefore, though there is no conclusive statement on the stability of quark matter (at least for the 2-flavor one), one important message from Figure~\ref{fig:result_1000} is that the combined information from pQCD and $\mbox{LQCD}_{\rm I}$ is approaching the answer to the question of quark matter stability.

In Figure~\ref{fig:result_1000}, we also show existing lowered bounds on the bag parameter from the Gell-Mann, Oakes, Renner (GMOR) relation~\cite{Gell-Mann:1968hlm} and the QCD sum rules, which we briefly mentioned in Section~\ref{sec:intro}. Note that the term $\sum_q m_q \bar{q} q$ in the trace anomaly defined in Eq.~\eqref{eq:trace_anomaly} is independent of the renormalization scale; the same is true for the sum of the two remaining ($\beta(g)$ and $\gamma_m(g)$) terms~\cite{Tarrach:1981bi}. 

For the contribution of $B_m \equiv \langle \sum_q m_q \bar{q} q \rangle$ to $B$, we have a reliable answer from the GMOR relation~\cite{Gell-Mann:1968hlm}, especially for the two light ($u$ and $d$) quarks
\begin{equation}
    \frac{f_\pi^2m_\pi^2}{4} = -m_\ell\langle\overline{\ell}\ell\rangle ~,
\end{equation}
where $f_\pi=130$~MeV and $m_\ell=(m_u+m_d)/2$, which in the $\overline{\rm MS}$ scheme at 2~GeV using the $SU(2)$ chiral perturbation theory gives $\langle\overline{\ell}\ell\rangle^{\overline{\rm MS}}(2~{\rm GeV})=-(272\pm2~{\rm MeV})^3$~\cite{Borsanyi:2012zv} and is consistent with the lattice results~\cite{McNeile:2012xh}. On the other hand, since $m_s$ is relatively heavy compared to $m_u$ and $m_d$, there is a greater disagreement among different measurements. For instance, Ref.~\cite{Harnett:2021zug} has summarized a couple of different measurements, from which we especially quote
\begin{equation}
    \frac{\langle\overline{s}s\rangle^{\overline{\rm MS}}(2~{\rm GeV})}{\langle\overline{\ell}\ell\rangle^{\overline{\rm MS}}(2~{\rm GeV})} = \begin{cases}
        1.08\pm0.16 &~[\text{Lattice~\cite{McNeile:2012xh}}] \\[1ex]
        0.66\pm0.1 &~[\text{Sum Rule Average~\cite{Narison:2004}}]
    \end{cases} ~.
\end{equation}

On the other hand, there is a large uncertainty in the second contribution of $B_G \equiv \langle \beta\,GG/2g + \sum_q m_q\,\gamma_m\,\bar{q} q\rangle$ to $B$. See for example Ref.~\cite{Narison:2018dcr} for an overview of various different measurements of the gluon condensate $\langle\alpha_s GG\rangle$. Some values of $\langle \alpha_s GG\rangle$ (not renormalization-scale-independent) obtained from the sum-rule approach are in conflict with the upper bound on $B$ extracted from the LQCD$_{\rm I}$ data in Section~\ref{sec:lattice}. As a result, we remain agnostic to this second contribution to $B$ and only assume $B_G \ge 0$. Accordingly, we have a trustable lower bound on $B$ 
\begin{equation}\label{eq:low_3-flavor}
    B \geq B_{{\rm min}, 2+1}\equiv -\frac{1}{4}\sum_{q=u,d,s}\langle m_q\overline{q}q\rangle^{\overline{\rm MS}}(2~{\rm GeV}) 
    = \begin{cases}
        (153\substack{+6.0 \\ -6.0}~{\rm MeV})^4 &~[\text{Lattice}] \\[1ex]
        (136\pm5.0~{\rm MeV})^4 &~[\text{Sum Rule Average}]
    \end{cases} ~.
\end{equation}
 
For the 2-flavor system, since the $\langle m_s \overline{s}s\rangle$ condensate may not be restored~\cite{Holdom:2017gdc}, one could only consider the contributions from the $u$- and $d$-quark chiral condensates to the bag parameter $B$. In this case, one would have a lower value for the lower bound
\begin{equation}\label{eq:low_2-flavor}
    B  \geq B_{{\rm min}, 2} \equiv -\frac{1}{4}\sum_{q=u,d}\langle m_q\overline{q}q\rangle^{\overline{\rm MS}}(2~{\rm GeV}) = (76.8\substack{+0.9 \\ -2.0}~{\rm MeV})^4 ~~[\text{GMOR}] ~.
\end{equation}

As can be seen from the left panel of Figure~\ref{fig:result_1000}, the lower bound on $B$ from Eq.~\eqref{eq:low_3-flavor} has entirely excluded the existence of the stable 2+1-flavor quark matter in the $X$ range of our interest. Given that in order for the 2+1-flavor quark matter to be stable, the trend of the stability constraint contour (the blue solid line) requires a smaller value of $B$ for a smaller value of $X$, one can draw the interesting conclusion that {\it there is no stable 2+1-flavor quark matter}. We also note that the $\langle m_s \bar{s} s\rangle$ condensate may not be restored in the isospin-dense matter. In this case, the upper bound on $B$ from LQCD$_{\rm I}$ can be treated as a more conservative bound for the 2+1-flavor quark matter. 

In the right panel, we compare the 2-flavor quark matter stability constraint contour (the orange solid line) with the LQCD$_{\rm I}$ constraints and the lower bound $B_{{\rm min}, 2}$ from the GMOR relation. One can see that there is only a narrow region of $X$ and $B$ that allows stable 2-flavor quark matter. It will be interesting to see if the future determination of $B$ falls into this allowed region.

Assuming more simulations on the LQCD side, one could try to make projected future sensitivity estimates. In general, there are two possibilities depending on the central value of the bag parameter. The first possibility is that one still cannot resolve or obtain a nonzero value for $B$. In this case, one can simply project the 90\% CL contours by uniformly scaling the bin-wise uncertainties by a factor of $1/N_{\rm scale}$, as we have shown in Figure~\ref{fig:lattice} for $N_{\rm scale}=\sqrt{10}$, while maintaining the same number of degrees of freedom $k$. In practice, the actual values of $k$ depend on the detailed future LQCD simulations and are to be determined. Here, to guide the discussion while anticipating a factor of 10 more simulated LQCD ensembles, we show the projected contours with $N_{\rm scale}=\sqrt{10}$ in red and green dot-dashed lines in Figure~\ref{fig:result_1000}. One can see that if the best-fit values of $X$ and $B$ do not shift dramatically, a conclusive statement for stable quark matter can be made, even though the actual value of $B$ is yet determined.

On the other hand, if $B$ turns out to be resolved with future LQCD simulations, there are three possible conclusions:
\begin{enumerate}
    \item If the best-fit point or benchmark has $X \leq 1.42$ (a dramatic change from the value based on the current LQCD data), then the quark matter stability remains inconclusive in the pQCD-trustable region. Some other approach is still needed to make the final determination.
    \item If the benchmark falls within the blue shaded region in the left panel or the orange shaded region in the right panel of Figure~\ref{fig:result_1000}, then the existence of any type of stable quark matter is excluded by the pQCD calculation.
    \item If the benchmark falls within the remaining white region in the right panel of Figure~\ref{fig:result_1000}, then it implies that the 2-flavor quark matter is more stable than ordinary nuclei based on the pQCD calculation.
\end{enumerate}

\section{General bounds on quark matter stability from thermodynamics } \label{sec:thermo}

In this section, we adopt a complementary approach to the main LQCD+pQCD framework of this work to explore the quark matter stability by applying the thermodynamic bounds on the quark matter equation of state (EOS) following the methodology outlined in Ref.~\cite{Komoltsev:2021jzg} (also considered in Ref.~\cite{Fujimoto:2023unl}). In this approach, one imposes a few basic thermodynamic requirements that can also provide a bound on the relation between the quark matter energy per baryon $\epsilon/\nB$ and the bag parameter $B$, and hence the stability of quark matter. The requirements are: 
1. the pressure $p(\muB)$ being a continuous function of $\muB$; 
2. the thermodynamic stability or concavity of the grand potential; 
3. the causality bound that requires the speed of sound be smaller than the speed of light;
and
4. the low-density reference point with $(\mu_L, p_L)$ and high-density reference point with $(\mu_H, p_H)$ ($\mu_L$ and $\mu_H$ are baryon chemical potentials) being connected by an integral relation. 
For the details of the framework, we refer to Appendix~\ref{sec:bound} and Appendix~\ref{sec:EOS}. 

The result produced by this general method is a robust but less stringent constraint on the relation between $\epsilon/\nB$ and $B$. Before we show the results, we first highlight the reasons:
\begin{enumerate}
    \item This method only requires a high-density reference point at a certain $\muB=\mu_H$ and is thus in principle immune to the uncertainty of pQCD calculations at lower densities as long as $\mu_H$ is chosen to be large enough to trust the pQCD calculation.
    \item The guiding principles are only the equilibrium conditions for a thermodynamic system and thus suffer from fewer uncertainties compared to the LQCD+pQCD framework.
    \item Since it considers all possible EOS's interpolating between the low- and high-density reference points while being agnostic to the fundamental QCD, it only provides a constraint on rather than a prediction for $\epsilon/\nB$.
\end{enumerate}
Therefore, the general thermodynamic approach can serve as a proxy to exploring the low-density parameter space where pQCD might be invalid or the LQCD data sensitivity can yet reach. For instance, the quark matter EOS may be obtained from studying the core properties of a hybrid neutron star. In the cases where the corresponding $p$ and $\muB$ are below the pQCD-trustable region, the general thermodynamic approach can be used to infer the stability of quark matter.

\begin{figure}[htb!]
    \centering
    \includegraphics[width=0.65\textwidth]{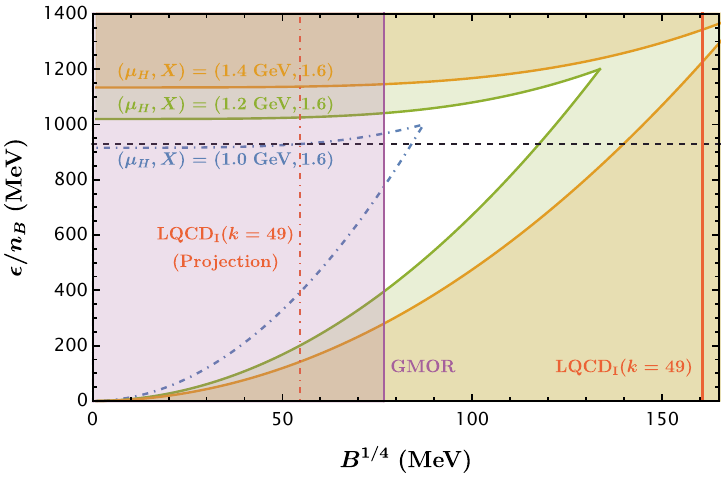}
    \caption{General thermodynamic bounds on the $B^{1/4}$--$(\epsilon/\nB)$ relation for different high-density reference benchmarks, based on the 2-flavor pQCD calculations. We also show the current (vertical red solid line) and projected (vertical red dot-dashed line) LQCD sensitivities for $X=1.6$, $\mu_{\rm start} = 1000$~MeV, and $k = 49$ (see Figure~\ref{fig:result_1000}). The horizontal black dashed line represents $\epsilon/\nB = 930$~MeV. The lower (GMOR) bound from Eq.~\eqref{eq:low_2-flavor} is also shown in the plot.
    }
    \label{fig:E-B-quark}
\end{figure}

In Figure~\ref{fig:E-B-quark}, we show such constraints on the $B^{1/4}$--$(\epsilon/\nB)$ parameter space derived from the general thermodynamic requirements using three example 2-flavor benchmarks with $(\mu_H, X) = (1.4~{\rm GeV}, 1.6)$, $(1.2~{\rm GeV}, 1.6)$, and $(1.0~{\rm GeV}, 1.6)$. One can see that the constraints are more stringent for a lower value of $\mu_H$. On the other hand, a lower value of $\mu_H$ implies a larger uncertainty in $p_H$ based on the pQCD calculation, so the derived constraints are less reliable. For the three benchmarks, the QCD gauge couplings are $\alpha_s(2 X \mu_H / 3) \approx 0.36, 0.41, 0.49$, respectively, and thus one could trust the constraints derived from the first two benchmarks (green and orange shaded regions) but not (at least not completely) the last one. For illustration, we also show the constraining contour in terms of a blue dot-dashed line for the third benchmark. The intersections between the constraining contours and the black dashed line ($\epsilon/\nB = 930~\mbox{MeV}$) are $B^{1/4} = 120$~MeV (140~MeV) for the green (orange) curves. This means that for a bag parameter above these values, the existence of stable 2-flavor quark matter is inconsistent with the general requirements of thermodynamic equilibrium derived with the two corresponding high-density reference points. On the other hand, both stable and unstable quark matter could exist in the white allowed region, so no conclusive answers can be obtained from this more general approach.

For comparison, we also show the current and projected LQCD sensitivities on $B$ with $X=1.6$, $\mu_{\rm start} = 1000$~MeV, and $k = 49$ on this plot (see Figure~\ref{fig:result_1000}), as well as the lower bound from Eq.~\eqref{eq:low_2-flavor}. Both approaches provide comparable constraints on $B$, although the LQCD+pQCD framework can be improved with future LQCD studies.

\section{Discussion and conclusions}\label{sec:conclusions}

As we briefly discussed in Section~\ref{sec:combine}, improved LQCD studies of the isospin-dense matter system are warranted to be more sensitive to the bag parameter or even to measure its value. On the other hand, to more precisely determine the CS condensate contribution to the EOS of isospin-dense matter, an extension of the calculation to a higher order in $\alpha_s$ as well as more general solutions to the gap equations are called for. This could possibly relax the $\mu > m_{\rm D} > \Delta$ condition required for the validity of the approximate expressions (see Appendix~\ref{sec:CFL-CS}) and thus extend the optimal range of the goodness-of-fit test.

Another important question that we touched on in Section~\ref{sec:pQCD} is the competition between the 2+1- and 2-flavor quark matter states as the global ground state (see the right panel of Figure~\ref{fig:pQCD_constraint}). While whether the stable quark matter state should be strange or strangeless has long been debated, the LQCD+pQCD framework of this work implies that depending on the $X$ and $B$ values of the specific system, either of them can be the global ground state. A more precise determination of $X$ and $B$ can in principle answer the question of the 2+1- and 2-flavor quark matter issue.

On the other hand, as we have shown previously, the LQCD+pQCD analysis is limited to $X>1.42$. For a system below that threshold, other measures are required to address the question of quark matter stability. For instance, one could study the quark core properties of a hybrid neutron star to infer the quark matter EOS at a high density that is below the pQCD-controlled range, from which one could extrapolate the relation to a low-density setting (using general thermodynamic conditions) and infer the stability of the quark matter. Alternatively, one may directly search for the radioactive signatures of small quark matter or quark nuggets produced from a neutron star merger (see Ref.~\cite{Bai:2024muo} for instance).

Parallel to the LQCD studies at finite density, LQCD simulations at finite temperature are another way to potentially measure the bag parameter and determine the stability of quark matter together with pQCD calculations. For instance, Ref.~\cite{Petreczky:2009at} has reviewed the lattice data from Refs.~\cite{Bernard:2006nj,Cheng:2007jq,Bazavov:2009zn} and performed a fit on the trace anomaly $\epsilon-3p$ using the $B$ parameter and the leading order pQCD contribution to obtain $B^{1/4} \simeq 200~{\rm MeV}$. One could extend this study by adopting higher-order pQCD-calculated results and treating $X$ as a model parameter, similar to what we have done in this study, to infer the stability of quark matter. More recently, Refs.~\cite{HotQCD:2014kol,Bazavov:2017dsy,Weber:2018bam} have performed simulations at even higher temperatures and analyzed the corresponding trace anomalies, which can be used to perform a more robust fit on $B$. By combining the LQCD studies at finite density and finite temperature, a more precise understanding of the bag parameter and thus the stability of quark matter could potentially be obtained.

In conclusion, we have first classified two possible relations between the hadron and quark phases of the baryon matter, which can be used to classify the existence scenarios of ordinary neutron stars, quark stars, and hybrid stars. By combining the pQCD+CS calculations and LQCD simulations for isospin-dense matter, we have inferred the sensitivity on the bag parameter $B$ and used it to constrain the possible existence of stable 2+1-flavor and/or 2-flavor quark matter. Promoting the ratio of the renormalization scale over the weighted sum of the quark chemical potentials to a model parameter $X$, the current LQCD calculations for isospin-dense matter are used to constrain the 2-dimensional parameter space of $X$ and $B^{1/4}$. The 90\%~CL region for $X$ and $B^{1/4}$ overlaps with the pQCD-trustable region, which hints that the quark matter stability issue could be addressed by the LQCD+pQCD framework. Furthermore, an upper bound on $B^{1/4} \lesssim 160$~MeV has been obtained, which is just around the corner from addressing the stability of quark matter. On the other hand, the $\langle m_q\overline{q}q\rangle$ condensates provide a minimum contribution to or a lower bound on $B$, which excludes the possible existence of the stable 2+1-flavor quark matter while leaving a tight parameter space for the stable 2-flavor quark matter up to two caveats: the still unknown $\mathcal{O}(\alpha_s^3)$ (even higher-order) corrections to $p_{\rm pQCD}$ and the possible partial restoration of the quark and gluon condensates. For comparison, we have also worked out the more general thermodynamic bounds on the quark matter energy-per-baryon and $B$ and found that they provide a weaker but complementary constraint.

\vspace{1cm}
\subsubsection*{Acknowledgments}
We thank Ryan Abbott and Mrunal Korwar for useful discussion. The work is supported by the U.S. Department of Energy under the contract DE-SC-0017647. TKC is also supported by the Ministry of Education, Taiwan, under the Government Scholarship to Study Abroad.

\begin{appendix}
\section{pQCD formulas for quark matter properties}\label{sec:quark}

In this section, we summarize the formulas used to calculate the quark matter properties up to $\mathcal{O}(\alpha_s^2)$ in Ref.~\cite{Kurkela:2009gj}. Assuming that there is only one massive quark flavor, the total number of quark flavors is $N_f=N_l+1$, where $N_l = 2$ denotes the number of massless quark flavors. Treating the strange quark as the only massive flavor, we denote its renormalized mass and chemical potential as $m$ and $\mu$, respectively. The gauge coupling and quark mass are functions of the renormalization scale $\mu_R$ as 
\beqa
    \alpha_s(\mu_R) &=& \frac{4\pi}{\beta_0L}\left(1-\frac{2\beta_1}{\beta_0^2}\frac{\ln L}{L}\right) ,~\qquad \mbox{with}\quad L = \ln\left(\mu_R^2/\Lambda_{\overline{\rm MS}}^2\right) ~, \\
    m(\mu_R) &=& m(2\,{\rm GeV})\left(\frac{\alpha_s(\mu_R)}{\alpha_s(2\,{\rm GeV})}\right)^{\gamma_0/\beta_0}\times\frac{1+A_1\frac{\alpha_s(\mu_R)}{\pi}+\frac{A_1^2+A_2}{2}\left(\frac{\alpha_s(\mu_R)}{\pi}\right)^2}{1+A_1\frac{\alpha_s(2\,{\rm GeV})}{\pi}+\frac{A_1^2+A_2}{2}\left(\frac{\alpha_s(2\,{\rm GeV})}{\pi}\right)^2} ~,
\eeqa
where we choose $\Lambda_{\overline{\rm MS}}=378$~MeV and
\begin{equation}
    A_1 = -\frac{\beta_1\gamma_0}{2\beta_0^2} + \frac{\gamma_1}{4\beta_0} ,\qquad A_2 = \frac{\gamma_0}{4\beta_0^2}\left(\frac{\beta_1^2}{\beta_0}-\beta_2\right) - \frac{\beta_1\gamma_1}{8\beta_0^2} + \frac{\gamma_2}{16\beta_0} ~.
\end{equation}
The $SU(3)_C$ group theory factors as well as the $\beta$ and $\gamma$ functions are summarized as follows:
\beqa
&&    d_A = N_c^2 - 1 ,~ \qquad C_A = N_c ,~ \qquad C_F = \frac{N_c^2-1}{2N_c} ~, \nonumber \\
&&    \beta_0 = \frac{11C_A-2N_f}{3} ,\qquad \qquad \beta_1 = \frac{17}{3}C_A^2 - C_FN_f - \frac{5}{3}C_AN_f  ~,  \nonumber \\
&&    \beta_2 = \frac{2857}{216}C_A^3 + \frac{1}{4}C_F^2N_f - \frac{205}{72}C_AC_FN_f - \frac{1415}{216}C_A^2N_f + \frac{11}{36}C_FN_f^2 + \frac{79}{216}C_AN_f^2 ~, \nonumber \\
&&    \gamma_0 = 3C_F , \qquad \qquad \qquad \gamma_1 = C_F\left(\frac{97}{6}C_A+\frac{3}{2}C_F-\frac{5}{3}N_f\right) ~,  \nonumber \\
&&    \gamma_2 = C_F\Bigg\{\frac{129}{2}C_F^2-\frac{129}{4}C_FC_A+\frac{11413}{108}C_A^2+C_FN_f\left[-23+24\zeta(3)\right] \nonumber \\
&&   \hspace{1.5cm} + \, C_AN_f\left[-\frac{278}{27}-24\zeta(3)\right]-\frac{35}{27}N_f^2\Bigg\} ~. \label{eq:QCD:group}
\eeqa
We also define the following abbreviations, which will be used later:
\begin{equation}
    u = \sqrt{\mu^2-m^2} \,,~ \uhat = u/\mu \,,~ \mhat = m/\mu \,,~ \vhat=1-\mhat \,,~ z = \uhat-\mhat^2\ln\left(\frac{1+\uhat}{\mhat}\right) ~.
\end{equation}

The grand potential of the quark matter $\Omega$ can be decomposed into four components: the massless quark contribution $\Omega^{m=0}$,
\begin{equation}\label{eq:Omega:m0}
\begin{aligned}
    \frac{\Omega^{m=0}}{V} &= -\frac{1}{4\pi^2}\sum_{i=1}^{N_l}\mu_i^4\,\Bigg\{\frac{N_c}{3}-d_A\frac{g^2(\mu_R)}{(4\pi)^2} \\
    & \hspace{0.5cm} + d_A\left[\frac{4}{3}\left(N_f-\frac{11C_A}{2}\right)\ln\frac{\mu_R}{2\mu_i}-\frac{142}{9}C_A+\frac{17}{2}C_F+\frac{22}{9}N_f\right]\frac{g^4(\mu_R)}{(4\pi)^4} \Bigg\} ~,
\end{aligned}
\end{equation}
the massive quark contribution $\Omega^m$,
\begin{align}\label{eq:Omega:m}
    \frac{\Omega^m}{V} &= \mathcal{M}_1 + \frac{g^2(\mu_R)}{(4\pi)^2}\mathcal{M}_2 + \frac{g^4(\mu_R)}{(4\pi)^4}\mathcal{M}_3 ~, \\
    \mathcal{M}_1 &= -\frac{N_c\,\mu^4}{24\pi^2}\left( 2\uhat^3-3z\mhat^2 \right) ~, \\
    \mathcal{M}_2 &= -\frac{d_A\,\mu^4}{4\pi^2}\left( -6z\mhat^2\ln\frac{\mu_R}{m} + 2\uhat^4 - 4z\mhat^2 - 3z^2 \right) ~, \\
    \mathcal{M}_3 &= -\frac{d_A\,\mu^4}{2\pi^2}\Bigg\{
        -\mhat^2\left[(11C_A-2N_f)z+18C_F(2z-\uhat)\right]\left(\ln\frac{\mu_R}{m}\right)^2 \nonumber \\
        &  \hspace{0.5cm} + \frac{1}{3}\Bigg[ C_A\left(22\uhat^4-\frac{185}{2}z\mhat^2-33z^2\right)+\frac{9C_F}{2}\left(16\mhat^2\uhat(1-\uhat)-3(7\mhat^2-8\uhat)z-24z^2\right) \nonumber \\
        &  \hspace{0.5cm} -N_f\left(4\uhat^4-13z\mhat^2-6z^2\right)\Bigg]\ln\frac{\mu_R}{m} + C_A\left(-\frac{11}{3}\ln\frac{\mhat}{2}-\frac{71}{9}+\mathcal{G}_1(\mhat)\right)\nonumber \\
        &  \hspace{0.5cm} + C_F\left(\frac{17}{4}+\mathcal{G}_2(\mhat)\right)  + N_f\left(\frac{2}{3}\ln\frac{\mhat}{2}+\frac{11}{9}+\mathcal{G}_3(\mhat)\right) + \mathcal{G}_4(\mhat)
    \Bigg\} ~, \\
    \mathcal{G}_1(\mhat) &= 32\pi^4\mhat^2\Big[ -0.01863+0.02038\mhat^2-0.03900\mhat^2\ln\mhat \nonumber \\
    & \hspace{0.5cm} +0.02581\mhat^2\ln^2\mhat-0.03153\mhat^2\ln^3\mhat+0.01151\mhat^2\ln^4\mhat \Big] ~, \\
    \mathcal{G}_2(\mhat) &= 32\pi^4\mhat^2\Big[ -0.1998-0.04797\ln\mhat+0.1988\mhat^2-0.3569\mhat^2\ln\mhat \nonumber \\
    & \hspace{0.5cm} +0.3043\mhat^2\ln^2\mhat - 0.1611\mhat^2\ln^3\mhat + 0.09791\mhat^2\ln^4\mhat \Big] ~, \\
    \mathcal{G}_3(\mhat) &= 32\pi^4\mhat^2\Big[ -0.05741-0.02679\ln\mhat-0.002828\ln^2\mhat+0.05716\mhat^2 \nonumber \\
    & \hspace{0.5cm} -0.08777\mhat^2\ln\mhat + 0.0666\mhat^2\ln^2\mhat-0.02381\mhat^2\ln^3\mhat + 0.01384\mhat^2\ln^4\mhat \Big] ~, \\
    \mathcal{G}_4(\mhat) &= 32\pi^4\mhat^2\bigg[ 0.07823+0.0388\ln\mhat+0.004873\ln^2\mhat -0.07822\mhat^2 + 0.1183\mhat^2\ln\mhat \nonumber \\
    &  \hspace{0.5cm} - 0.08755\mhat^2\ln^2\mhat + 0.03293\mhat^2\ln^3\mhat - 0.01644\mhat^2\ln^4\mhat \bigg] ~,
\end{align}
the vacuum-matter cross term $\Omega^x_{\rm VM}$,
\begin{align}\label{eq:Omega:VM}
    \frac{\Omega^x_{\rm VM}}{V} &= d_A\frac{g^4(\mu_R)}{(4\pi)^2}\frac{m^4}{12\pi^4}\sum_{i=1}^{N_l} I_x\left(\frac{\mu_i}{m+\mu_i}\right) ~, \\
    I_x(t) &\simeq -3t^4(1-\ln t)\Bigg\{\frac{0.83}{(1-t)^2}+\frac{0.06}{1-t}-0.056 \nonumber \\
    & \hspace{0.5cm} + \frac{\ln(1-t)}{t(1-t)^2}\left[1.005-0.272t(1-t)+0.154t(1-t)^2\right] \Bigg\} ~,
\end{align}
and the plasmon ring sum $\Omega_{\rm ring}$,
\begin{equation}\label{eq:Omega:ring}
\begin{aligned}
    \frac{\Omega_{\rm ring}}{V} &= \frac{d_A\,g^4}{512\pi^6}\Bigg\{
        \vec{\mu}^4\Bigg[
            2\ln\left(\frac{g}{4\pi}\right) - \frac{1}{2} + \frac{1}{2}\left(-\frac{19}{3}+\frac{2\pi^2}{3}+\frac{I_{15}(\vec{\mu})}{\vec{\mu}^4}+\frac{16}{3}(1-\ln2)\ln2+I_{16}(\mhat,\vec{\hat{\mu}}^2\right)
        \Bigg] \\
        & \hspace{0.5cm} + 2\mu^2\sum_{i=1}^{N_l}\mu_i^2\Bigg[
            I_{14}\left(2\ln\left(\frac{g}{4\pi}\right)-\frac{1}{2}\right) + \frac{1}{2}\left(I_{17}(\mhat,\hat{\mu}_i)+\frac{16}{3}(1-\ln2)\ln2 I_{18} + I_{19}\left(\mhat,\vec{\hat{\mu}}^2\right)\right)
        \Bigg] \\
        &  \hspace{0.5cm} + \mu^4\Bigg[
            I_{13}\left(2\ln\left(\frac{g}{4\pi}\right)-\frac{1}{2}\right)+\frac{1}{2}\left(I_{20}+\frac{16}{3}(1-\ln2)\ln2 I_{21}+I_{22}\left(\mhat,\vec{\hat{\mu}}^2\right)\right)
        \Bigg]
    \Bigg\} ~,
\end{aligned}
\end{equation}
where $\vec{\mu}=(\mu_1,\mu_2,\cdots,\mu_{N_l})$ and $\vec{\hat{\mu}}=\vec{\mu}/\mu$. Note that $\mu$ is the strange quark chemical potential, but not $\vert\vec{\mu}\vert$. The approximations of the integral functions ($I_{13-22}$) are listed in Appendix~\ref{sec:integral}. Overall, one finds that
\begin{equation}
    \Omega = \Omega^{m=0} + \Omega^m + \Omega^x_{\rm VM} + \Omega_{\rm ring} ~.
\end{equation}
In the special case where all flavors are considered massless, one has
\begin{equation}
    \Omega_{\rm massless} = \Omega^{m=0} + \Omega_{\rm ring}(m=0) ~.
\end{equation}

For the 2- and 2+1-flavor systems, $\beta$-equilibrium and charge neutrality require that
\begin{align}
    \text{2-flavor}&:~\mu_u=\mu_d-\mu_e \,,~ \frac{2}{3}n_u - \frac{1}{3}n_d - n_e = 0 ~, \\
    \text{2+1-flavor}&:~\mu_d=\mu_s \,,~ \mu_u=\mu_d-\mu_e \,,~ \frac{2}{3}n_u - \frac{1}{3}n_d-\frac{1}{3}n_s - n_e = 0 ~,
\end{align}
where $n_e=\mu_e^3/(3\pi^2)$ (ignoring electron mass), which implies that $\Omega_e/V=-\mu_e^4/(12\pi^2)$, and $n_q=-(1/V)(\partial\Omega/\partial\mu_q)$. With these conditions, there will only be one independent chemical potential, which we choose to be $\mu\equiv \mu_d$.

With these formulas, one would find a particular chemical potential $\mu=\mu_0(X)$, where $X$ is defined in Eq.~\eqref{eq:X-param}, below which some of the number densities become negative. Because of the charge neutrality condition, the number densities of the different particles are correlated. This will force $n_u$, $n_d$, and $n_e$ to simultaneously become zero at $\mu=\mu_0(X)$, while $n_s$ could become zero at a higher chemical potential because of the finite strange quark mass. Correspondingly, $\muB(\mu=\mu_0(X))=0$. This signals the breakdown of the calculations in that regime, and thus one should terminate the calculation at this point and set $p_{\rm pQCD}(\mu=\mu_0,X)=-\Omega(\mu=\mu_0,X)/V=0$, or equivalently
\begin{equation}
    p_{\rm pQCD}(\mu,X) = -\frac{1}{V}\left[\Omega(\mu,X)-\Omega(\mu=\mu_0,X)\right] ~.
\end{equation}

\begin{figure}[ht!]
    \centering
    \includegraphics[width=0.6\linewidth]{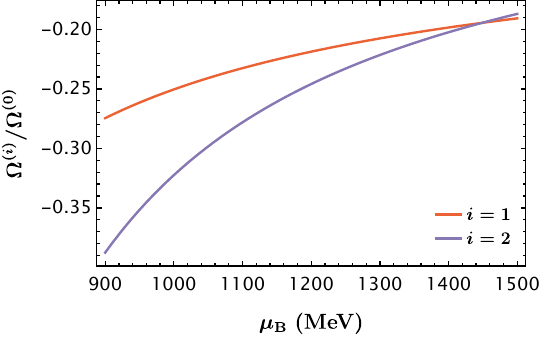}
    \caption{The ratios of the $\mathcal{O}[(\frac{\alpha_s}{\pi})^1]$ ($\Omega^{(1)}$) and $\mathcal{O}[(\frac{\alpha_s}{\pi})^2]$ ($\Omega^{(2)}$) contributions to the leading-order term $\mathcal{O}[(\frac{\alpha_s}{\pi})^0]$ ($\Omega^{(0)}$) in Eq.~\eqref{eq:Omega:m0} for $X=2$ and neglecting quark masses.}
    \label{fig:Omega_ratio}
\end{figure}

\begin{figure}[ht!]
    \centering
    \includegraphics[width=0.49\linewidth]{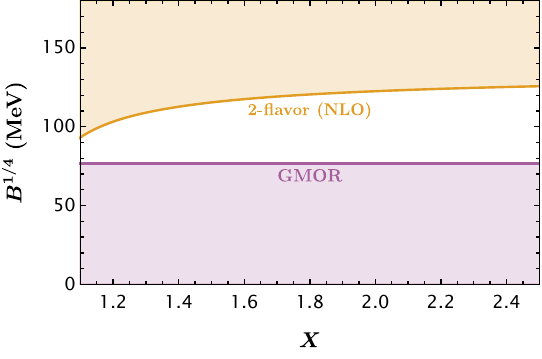}
    \includegraphics[width=0.49\linewidth]{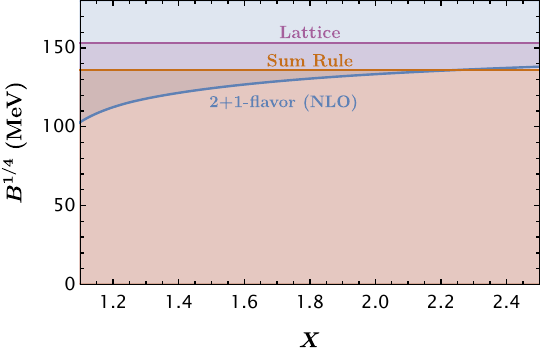}
    \caption{Similar plots to Figure~\ref{fig:result_1000} but with pQCD only up to $\mathcal{O}(\alpha_s^1)$.}
    \label{fig:result_NLO}
\end{figure}

To investigate the perturbativity of the pQCD calculations, we assume a simple massless three-flavor scenario (in which $\muB=3\mu$) and show in Figure~\ref{fig:Omega_ratio} the ratios of the $\mathcal{O}[(\frac{\alpha_s}{\pi})^1]$ ($\Omega^{(1)}$) and $\mathcal{O}[(\frac{\alpha_s}{\pi})^2]$ ($\Omega^{(2)}$) terms to the $\mathcal{O}[(\frac{\alpha_s}{\pi})^0]$ term ($\Omega^{(0)}$) in Eq.~\eqref{eq:Omega:m0} for $X=2$ (and, as mentioned in Section~\ref{sec:pQCD}, $(\frac{\alpha}{\pi})<1$ throughout the $\muB$ range of our interest), from which one can see that the individual ratios are both justifiably small. One might question about the behavior of the $\mathcal{O}[(\frac{\alpha_s}{\pi})^3]$ correction, which, if in the extreme limit takes up the positive sign with a magnitude comparable to the $\mathcal{O}[(\frac{\alpha_s}{\pi})^2]$ correction, can possibly reverse our conclusion about the stability of the 2+1-flavor quark matter (see Section~\ref{sec:combine}). To answer this question, we perform a similar analysis up to $\mathcal{O}[(\frac{\alpha_s}{\pi})^1]$ and show two similar plots to Figure~\ref{fig:result_1000} in Figure~\ref{fig:result_NLO}, from which one can see that the 2+1-flavor quark matter is still well excluded if we take the (more reliable) lattice result for the $s$-quark condensate, while for the sum rule average value, the conclusion still holds up to $X\sim2.2$, which already exceeds the usual maximum value of the renormalization scale parameter $X=2$ used in the literature. Although we cannot completely exclude the possibility of a quark matter system taking up $X>2$ or that the $\mathcal{O}[(\frac{\alpha_s}{\pi})^3]$ correction being large enough to invalidate the analysis of this study, we think they are unlikely to happen according to the partial results presented in Ref.~\cite{Gorda:2023mkk} (which also shows the increasing trend of $\mu_0$ at larger $X$ values that will further solidify the exclusion of stable 2+1-flavor quark matter) and can only be answered in future studies.

\begin{figure}[htb!]
    \centering
    \includegraphics[width=0.6\linewidth]{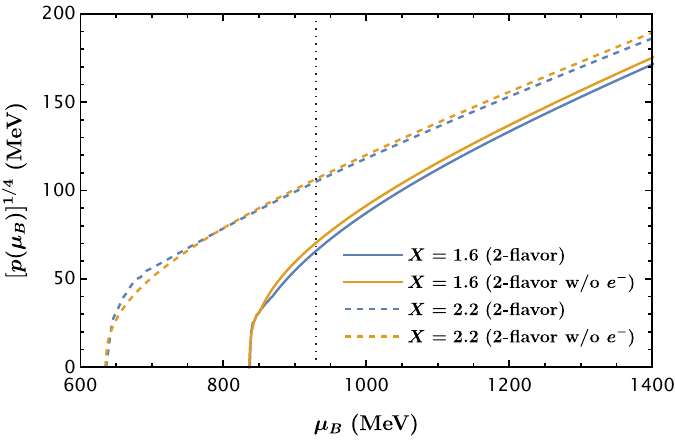}
    \caption{
    The $[p(\muB)]^{1/4}$ curves of the 2-flavor $X=1.6,2.2$ scenarios with and without electrons. For the scenarios without electrons, charge neutrality is satisfied solely by the $u$- and $d$-quark compositions or $2 n_u = n_d$.
    \label{fig:QCD_comp}
    }
\end{figure}

Before concluding this section, we remark that one can in principle also consider the scenarios without the $\beta$-equilibrium requirement or electrons and thus charge neutrality is satisfied solely by the different quark compositions. As pointed out in Ref.~\cite{Kurkela:2009gj}, electrons usually play a minor role in the overall thermodynamic properties of the system. For illustration, we show the $[p(\muB)]^{1/4}$ curves of the 2-flavor $X=1.6,2.2$ scenarios with and without electrons in Figure~\ref{fig:QCD_comp}, from which one can see that the two scenarios produce similar predictions.

\section{Integral approximations}\label{sec:integral}

We list in this section the integral approximations used in the pQCD calculations in Appendix~\ref{sec:quark}~\cite{Kurkela:2009gj}.
\begin{align}
    I_{13}(\mhat) &\simeq \frac{8}{3}(1-\ln2)\uhat^3 + 0.318\uhat^6 -0.137\uhat^7 ~, \\
    I_{14}(\mhat) &\simeq 1.99\vhat - 0.99\vhat^2 + \ln\vhat\left(-0.27\vhat+0.26\vhat^2-0.119\vhat^2\ln\vhat\right) ~, \\
    I_{15}(\vec{\mu}) &\simeq \frac{16}{3}\ln2\sum_i\mu_i^4 - 2\vec{\mu}^2\sum_i\mu_i^2\ln\frac{\mu_i^2}{\vec{\mu}^2} + \frac{2}{3}\sum_{i=1}^{N_l}\sum_{j>i}^{N_l} \Bigg[
        (\mu_i-\mu_j)^4\ln\frac{\vert\mu_i^2-\mu_j^2\vert}{\mu_i\mu_j} \nonumber \\
        &+ 4\mu_i\mu_j(\mu_i^2+\mu_j^2)\ln\frac{(\mu_i+\mu_j)^2}{\mu_i\mu_j}-(\mu_i^4-\mu_j^4)\ln\frac{\mu_i}{\mu_j}
    \Bigg] ~, \\
    I_{16}(\mhat,\vec{\hat{\mu}}^2) &= -0.8564+2\ln\left(\frac{\vec{\hat{\mu}}^4+2\vec{\hat{\mu}}^2I_{14}+I_{14}I_{16h1}}{\vec{\hat{\mu}}^4+\vec{\hat{\mu}}^2I_{14}}\right) ~, \\
    I_{16h1}(\mhat) &\simeq 4.629\vhat - 3.629\vhat^2 + \ln\vhat\left( 0.6064\vhat+2.021\vhat^2-0.566\vhat^2\ln\vhat \right) ~, \\
    I_{17}(\mhat,\hat{\mu}_i) &= \frac{I_{14}}{6\hat{\mu}_i^2}\Bigg[
        (1-\hat{\mu}_i)^4\ln\left(1-\frac{1}{\hat{\mu}_i}\right)^2
        +(1+\hat{\mu}_i)^4\ln\left(1+\frac{1}{\hat{\mu}_i}\right)^2
        -2\ln\frac{1}{\hat{\mu}_i^2}
    \Bigg] + I_{17h1} \nonumber \\
    & +I_{17h2}\frac{\hat{\mu}_i}{\hat{\mu}_i+1} + I_{17h3}\frac{1}{\hat{\mu}_i^2+1}+I_{17h4}\frac{\hat{\mu}_i}{3.38\sqrt{1-\mhat}+7.96(1-\mhat)+\hat{\mu}_i^3} ~, \\
    I_{17h1}(\mhat) &\simeq -7.5439\vhat + 1.1422\uhat + 6.639\uhat^5 - 8.8809\uhat^3\ln(1+\mhat) \nonumber \\
    & + \ln\vhat\left(-2.005\vhat+5.865\vhat^2\ln\vhat-10.622\vhat^3\ln^2\vhat\right) ~, \\
    I_{17h2}(\mhat) &\simeq -8.57\vhat + 3.09\uhat + 9.654\uhat^2 + 10.86\uhat^3 - 12.445\uhat^4 \nonumber \\
    & +\uhat\ln\vhat(0.374+8.73\uhat^2)-\frac{7}{3}I_{14}-I_{17h1} ~, \\
    I_{17h3}(\mhat) &\simeq 4.804\vhat - 2.224\vhat^2 + \ln\vhat(-0.411\vhat+0.056\vhat^2-0.0718\vhat^2\ln\vhat) - \frac{7}{3}I_{14}-I_{17h1} ~, \\
    I_{17h4}(\mhat) &\simeq -8.085\vhat^3+6.646\vhat^5-0.586\vhat^{12}+0.3465\mhat\uhat \nonumber \\
    & +1.359\uhat^2 - 3.77\uhat^3 + 5.395\uhat^4 - 0.9525\uhat^5 ~, \\
    I_{18}(\mhat) &\simeq 2.3\vhat-1.3\vhat^2+\ln\vhat\left(-0.3477\vhat+0.6484\vhat^2-0.241\vhat^2\ln\vhat\right) ~, \\
    I_{19}(\mhat,\vec{\hat{\mu}}^2) &= I_{19h1} + I_{14}\left[ 
        2\ln\vec{\hat{\mu}}^2+2\ln\left(
            \frac{I_{14}\vec{\hat{\mu}}^4+2I_{13}\vec{\hat{\mu}}^2+I_{14}I_{19h2}}{I_{14}\vec{\hat{\mu}}^4+I_{13}\vec{\hat{\mu}}^2}
        \right)
    \right] ~, \\
    I_{19h1}(\mhat) &\simeq -3.6643\vhat + 2.808\vhat^2 + \ln\vhat(-0.7305\vhat-1.22\vhat^2) \nonumber \\
    & + \ln^2\vhat(-0.0437\vhat+0.312\vhat^2-0.0237\vhat^2\ln\vhat) ~, \\
    I_{19h2}(\mhat) &\simeq 0.104\vhat + 0.464\vhat^2 + 0.432\uhat^5 + \ln\vhat(-0.04\vhat-0.99\vhat^2+0.455\vhat^2\ln\vhat) ~, \\
    I_{20}(\mhat) &\simeq 3.84\vhat - 7.94\vhat^2 + 8.06\uhat^2 + \ln\vhat(3.81\vhat+9.7\vhat^2-7.42\vhat^2\ln\vhat) ~, \\
    I_{21}(\mhat) &\simeq -0.11\vhat + 1.11\vhat^2 + \ln\vhat(-0.059\vhat-2.07\vhat^2+0.4\vhat^2\ln\vhat) ~, \\
    I_{22}(\mhat,\vec{\hat{\mu}}^2) &= I_{22h1} + I_{13}\left[
        2\ln\vec{\hat{\mu}}^2 + 2\ln\left(
            \frac{\vec{\hat{\mu}}^4+\vec{\hat{\mu}}^2I_{22h2}+I_{22h3}}{\vec{\hat{\mu}}^4+0.5\vec{\hat{\mu}}^2I_{22h2}}
        \right)
    \right] ~, \\
    I_{22h1}(\mhat) &\simeq -0.8255 \uhat^2 - 0.03084\uhat^7 + 0.00161\vhat\ln\vhat ~, \\
    I_{22h2}(\mhat) &\simeq 1.313\uhat + 0.434\uhat^2 + 0.253\uhat^5 ~, \\
    I_{22h3}(\mhat) &\simeq 1.844\vhat - 0.844\vhat^2 + \ln\vhat(0.194\vhat-0.392\vhat^2-0.704\vhat^2\ln\vhat) ~.
\end{align}

\section{Formulas of color superconductivity}\label{sec:CFL-CS}
In addition to the perturbative contributions, it is long known that color superconductivity (CS) in the color antitriplet $qq$ channel (color-flavor locking (CFL)) could also arise in the finite-density quark matter, which effectively contributes an extra term to the quark matter EOS. Moreover, a similar color singlet $\overline{q}q$ phase specific to the isospin-dense matter could also arise at finite density. Refs.~\cite{Fujimoto:2023mvc,Fujimoto:2024pcd} have calculated the associated pressures of these condensates up to the NLO.
The final results are given by
\begin{equation}\label{eq:P_SC}
    p_{\rm CS} = d\frac{\mu^2}{4\pi^2}\Delta^2\left[1+\frac{2\pi}{3^{3/2}\sqrt{c_R}}\left(\frac{\alpha_s}{\pi}\right)^{1/2}+\mathcal{O}(\alpha_s)\right] ~,
\end{equation}
with the gap $\Delta$ given by
\begin{equation}\label{eq:gap}
    \ln\left(\frac{\Delta}{\mu}\right) = -\frac{\sqrt{3}\,\pi}{2\,\sqrt{c_R}}\left(\frac{\alpha_s}{\pi}\right)^{-1/2} - \frac{5}{2}\ln\left(N_f\frac{\alpha_s}{\pi}\right) + \ln\frac{2^{13/2}}{\pi} - \frac{\pi^2+4}{12\,c_R} - \zeta + \mathcal{O}(\alpha_s^{1/2}) ~,
\end{equation}
where $\zeta=\frac{1}{3}\ln2\,(0)$, $c_R=2/3\,(4/3)$, and $d=12\,(6)$ for the CFL\,(color singlet $\bar{q}q$) phases in baryon-dense\,(isospin-dense) matter, respectively. Because of the different $c_R$ factors, $p_{\rm CS}$ for the CFL channel is much more suppressed compared to the pQCD contributions, while $p_{\rm CS}$ for the color singlet $\bar{q}q$ channel plays a major role in the analysis of the isospin-dense matter (see Section~\ref{sec:lattice}).

\section{Additional details of the Lattice QCD data analysis}\label{sec:lattice_details}

\begin{figure}[ht!]
    \centering
    \includegraphics[width=0.46\linewidth]{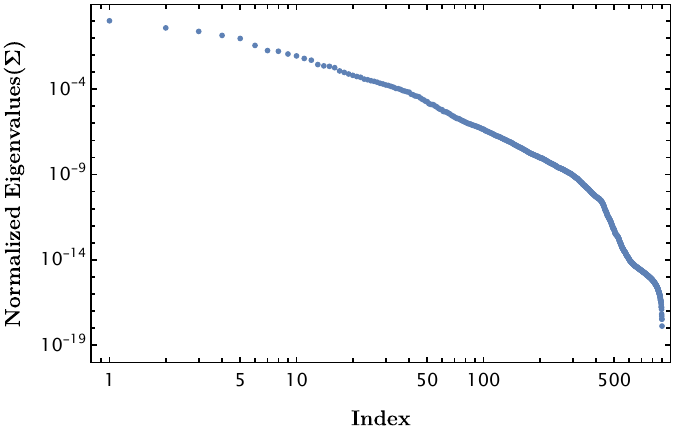} \hspace{8mm}
    \includegraphics[width=0.46\linewidth]{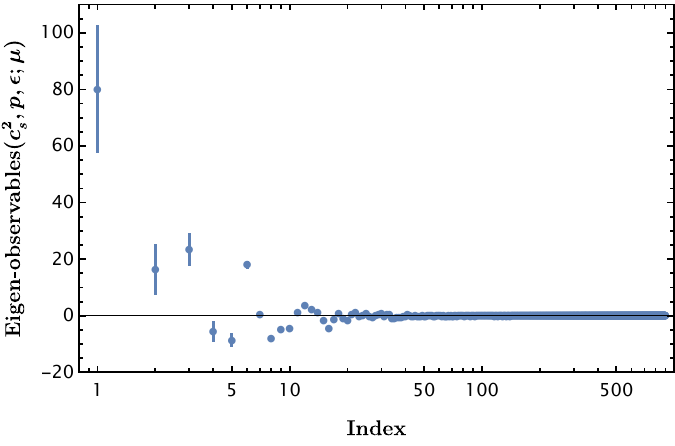}
    \caption{{\it Left panel:} The eigenvalues normalized by the maximum eigenvalue of the covariance matrix $\Sigma$. {\it Right panel:} The eigen-observables associated with the eigenvalues of $\Sigma$.
    }
    \label{fig:Sigma_eigen}
\end{figure}

We present in this section the details left out of Sections~\ref{sec:lattice} and \ref{sec:combine}.
The eigenvalues normalized by the maximum eigenvalue of the covariance matrix $\Sigma$ using the $\mbox{LQCD}_{\rm I}$ data from Ref.~\cite{Abbott:2024vhj} and the associated eigen-observables are presented in Figure~\ref{fig:Sigma_eigen}.

\begin{figure}[ht!]
    \centering
    \includegraphics[width=0.32\linewidth]{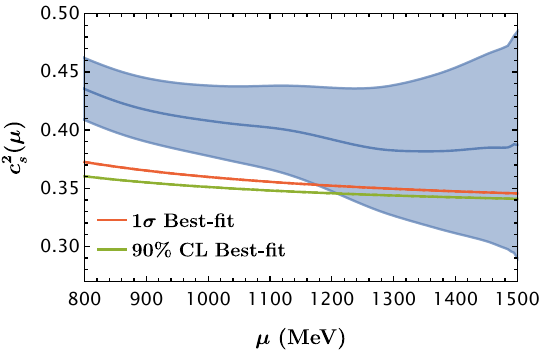} \hspace{1.0mm}
    \includegraphics[width=0.32\linewidth]{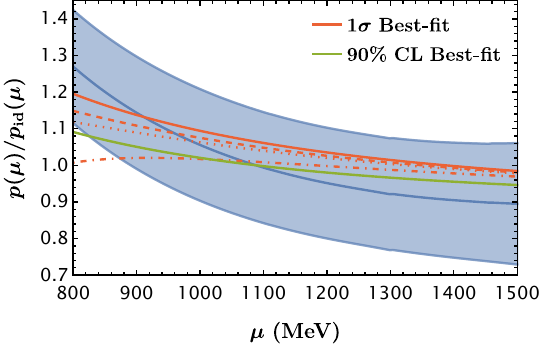} \hspace{1.0mm}
    \includegraphics[width=0.32\linewidth]{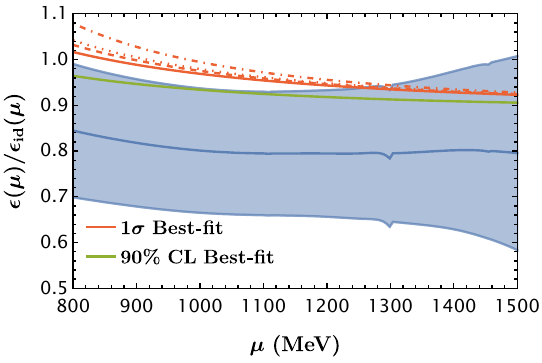}
    \caption{
    The $c_s^2$, $p/p_{\rm id}$, and $\epsilon/\epsilon_{\rm id}$ distributions of the best-fit points (solid lines) with $\mu_{\rm start}=800$~MeV and $k=30,35$, whose benchmark parameters are $(X,B^{1/4})=(1.30,0~{\rm MeV})$ and $(X,B^{1/4})=(1.58,0~{\rm MeV})$. We also show the distributions of the benchmarks $(X,B^{1/4})=(1.30,176~{\rm MeV})$ (red dashed line) which sits on the $90\%$\,CL contour around the $k=30$ best-fit point, $(X,B^{1/4})=(1.30,200~{\rm MeV})$ (red dotted line), and $(1.30,250~{\rm MeV})$ (red dot-dashed line). The red lines completely overlap with one another for $c_s^2$ since it does not depend on $B$.
    }
    \label{fig:fit_800}
\end{figure}

\begin{figure}[ht!]
    \centering
    \includegraphics[width=0.49\linewidth]{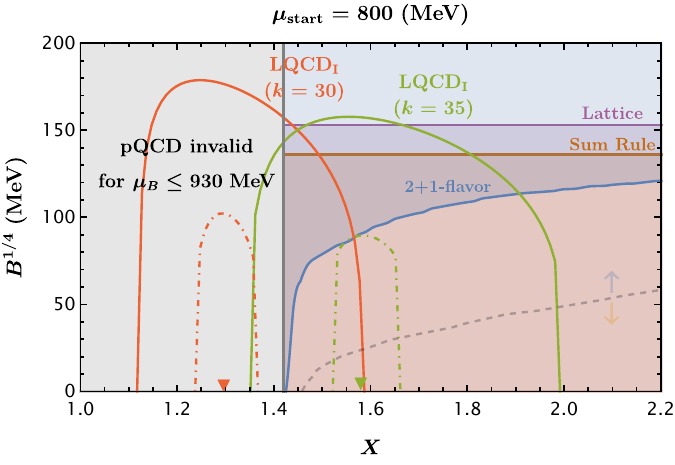}
    \includegraphics[width=0.49\linewidth]{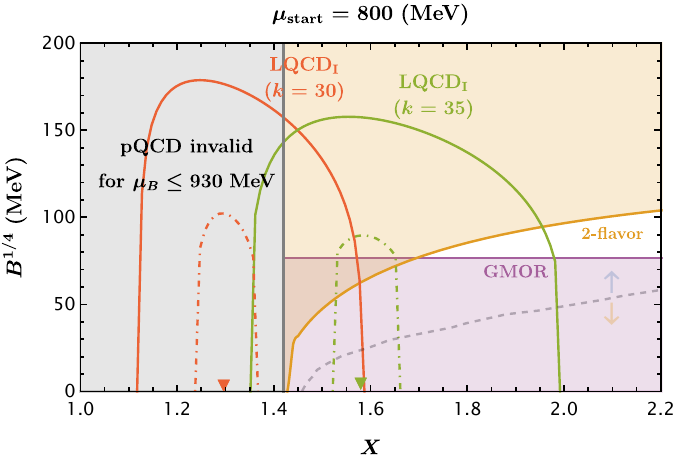}
    \caption{
    The $90\%$~CL ({\it left}) 2+1-flavor and ({\it right}) 2-flavor quark matter stability constraint for $\mu_{\rm start}=800$~MeV and $k=30,35$ (solid lines) as well as the corresponding future projection with $N_{\rm scale}=\sqrt{10}$ (dot-dashed lines). The lower bounds from Eqs.~\eqref{eq:low_3-flavor} and \eqref{eq:low_2-flavor} are also shown in the left and right panels, respectively.
    }
    \label{fig:result_800}
\end{figure}

We also present the results of the same analysis done in Sections~\ref{sec:lattice} and \ref{sec:combine} for $\mu_{\rm start}=800$~MeV in Figures~\ref{fig:fit_800} and \ref{fig:result_800}. The $1\sigma$ and $90\%$\,CL best-fit parameters are given by $(X,B^{1/4})=(1.30,0~{\rm MeV})$ and $(X,B^{1/4})=(1.58,0~{\rm MeV})$, while we also present the benchmarks of $(X,B^{1/4})=(1.30,176~{\rm MeV})$ (red dashed line) which sits on the $90\%$\,CL contour around the $k=30$ best-fit point, $(X,B^{1/4})=(1.30,200~{\rm MeV})$ (red dotted line), and $(1.30,250~{\rm MeV})$ (red dot-dashed line).

\section{Thermodynamic bounds on the QCD EOS}\label{sec:bound}

We review in this section the framework of the thermodynamic bounds on the QCD EOS presented in Ref.~\cite{Komoltsev:2021jzg}. Given two reference points (a low-density and a high-density one) defined by their baryon chemical potentials $\mu_{L,H}$, baryon number densities $n(\mu_{L,H})=n_{L,H}$, and pressures $p(\mu_{L,H})=p_{L,H}$, one can impose the following thermodynamic constraints on all the possible EOS's that interpolate between these two points:
\begin{enumerate}
    \item The pressure $p(\mu)$ should be a continuous function of $\mu$.
    \item Thermodynamic stability requires the concavity of the grand potential,
    \begin{equation}
        \partial_\mu^2\Omega(\mu)=-\partial_\mu^2 p(\mu)=-\partial_\mu n(\mu)\leq 0\to \partial_\mu n(\mu)\geq 0 ~.
    \end{equation}
    \item Causality requires that the speed of sound $c_s$ be less than the speed of light,
    \begin{equation}\label{eq:causality}
        \text{Causality constraint:} ~ c_s^{-2} = \frac{\mu}{n}\frac{\partial n}{\partial \mu} \geq 1 ~. \hspace{2.2cm}
    \end{equation}
    \item The EOS must simultaneously reach both $n_H$ and $p_H$ at $\mu=\mu_H$,
    \begin{equation}
        \text{Integral constraint:} ~ \int_{\mu_L}^{\mu_H} n(\mu)d\mu = p_H - p_L = \Delta p ~.
    \end{equation}
\end{enumerate}
For the details of these constraints, we refer to Ref.~\cite{Komoltsev:2021jzg}. We summarize the consequent bounds on $n(\mu)$ and $p(\mu,n)$ in the following. First, for $n(\mu)$:
\begin{equation}
    n_{\rm min}(\mu) \leq n(\mu) \leq n_{\rm max}(\mu) ~,
\end{equation}
\begin{align}
    n_{\rm min}(\mu) &= \begin{cases}
        \begin{aligned}
            &\frac{n_L}{\mu_L}\mu &,~\quad \mu_L \leq \mu \leq \mu_c \\[1ex]
            &\frac{\mu^3\,n_H - \mu\,\mu_H(\mu_H\,n_H-2\,\Delta p)}{(\mu^2-\mu_L^2)\,\mu_H} &,~\quad \mu_c < \mu \leq \mu_H
        \end{aligned}
    \end{cases} ~, \\[1ex]
    n_{\rm max}(\mu) &= \begin{cases}
        \begin{aligned}
            &\frac{\mu^3\,n_L-\mu\,\mu_L(\mu_L\,n_L+2\,\Delta p)}{(\mu^2-\mu_H^2)\,\mu_L} &,~\quad  \mu_L\leq \mu < \mu_c \\[1ex]
            &\frac{n_H}{\mu_H}\mu &,~\quad  \mu_c \leq \mu \leq \mu_H
        \end{aligned}
    \end{cases} ~,
\end{align}
where
\begin{equation}
    \mu_c = \sqrt{\frac{\mu_L\,\mu_H\,(\mu_H\,n_H-\mu_L\,n_L-2\,\Delta p)}{\mu_L\,n_H-\mu_H\,n_L}} ~.
\end{equation}
Then, for $p(\mu,n)$:
\begin{equation}
    p_{\rm min}(\mu,n) \leq p(\mu,n) \leq p_{\rm max}(\mu,n) ~,
\end{equation}
\begin{align}
    p_{\rm min}(\mu,n) &= p_L + \frac{\mu^2-\mu_L^2}{2\mu}n_{\rm min}(\mu) ~, \\[1ex]
    p_{\rm max}(\mu,n) &= \begin{cases}
        \begin{aligned}
            &p_L + \frac{\mu^2-\mu_L^2}{2\mu}n &,~\quad  n \leq n_c(\mu) \\[1ex]
            &p_H - \frac{\mu_H^2-\mu^2}{2\mu}n &,~ \quad n > n_c(\mu)
        \end{aligned} ~,
    \end{cases}
\end{align}
where
\begin{equation}
    n_c(\mu) = \frac{n_{\rm max}(\mu_L)}{\mu_L}\mu = \frac{n_{\rm min}(\mu_H)}{\mu_H}\mu ~.
\end{equation}
One can go on to derive $p_{\rm min}(\mu)$ and $p_{\rm max}(\mu)$, which can be found to be
\begin{align}
    p_{\rm min}(\mu) &= p_{\rm min}\big(\mu, n_{\rm min}(\mu)\big) ~, \\[1ex]
    p_{\rm max}(\mu) &= p_{\rm max}\big(\mu, n_c(\mu)\big) ~.
\end{align}

\section{Quark matter EOS and ground state energy}\label{sec:EOS}

\begin{figure}[ht!]
    \centering
    \includegraphics[width=0.46\textwidth]{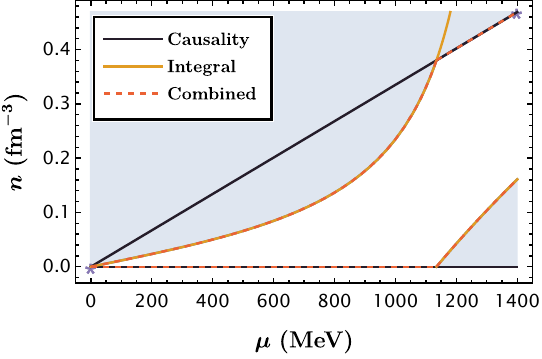}\hspace{3mm}
    \includegraphics[width=0.49\textwidth]{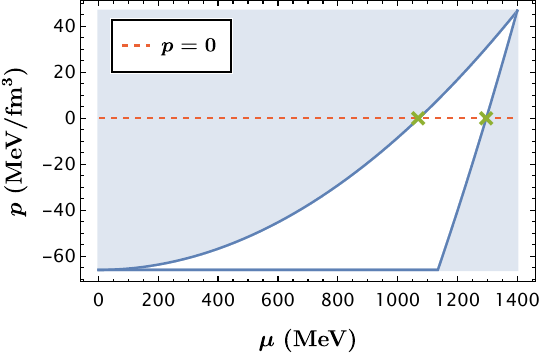}
    \caption{Thermodynamic bounds on ({\it left}) $n(\mu)$ and ({\it right}) $p(\mu)$ using the 2-flavor $X=1.6$ benchmark for the high-density reference point with $\mu_H=1400$~MeV and $B=(150~{\rm MeV})^4$. In the right panel, the green crosses are the intersections between the thermodynamics bounds and the $p=0$ line, which give the minimally and maximally allowed quark matter ground state energies per baryon ($E_{Q,{\rm min,max}}(B)=\mu_{Q,{\rm min,max}}(B)$), respectively.
    }
    \label{fig:EOS-quark}
\end{figure}

In this section, we study the thermodynamic constraints on the quark matter EOS. While in principle one can choose $\mu_L=\mu_0>0$ for the low-density reference point when constraining the quark matter EOS, the uncertainty of pQCD calculations at that point is very high and thus unreliable (see Sections~\ref{sec:classification} and \ref{sec:pQCD}). A more conservative choice is to choose $\mu_L=0$ with $n_L=0$ and will lead to a less stringent but more reliable bound, which we use in this study. To specify $p_L$, we need additional inputs about the deconfined vacuum in terms of the bag parameter $B$, with which we can set $p_L=-B$.
On the other hand, the high-density reference point can be obtained through pQCD calculations, and one should also shift $p_{{\rm pQCD}, H}\to p_H = p_{{\rm pQCD}, H}-B$ to include the nonperturbative corrections. We show one specific example of $p_L=-(150~{\rm MeV})^4$ and $(\mu_H,n_H,p_H)$ $=$ $(1400~{\rm MeV},0.469~{\rm fm^{-3}},112.6~{\rm MeV/fm^3})$, one benchmark of the 2-flavor $X=1.6$ scenario, in Figure~\ref{fig:EOS-quark}.

Using the Euler equation $\epsilon = -p + \mu n$, we can express the energy per baryon as $\epsilon/n \equiv E = -p/n + \mu$. The quark matter ground state corresponds to the thermodynamic configuration with the minimum $E(\mu=\mu_Q)$, which can be solved through
\begin{equation}
    \left.\frac{\partial E}{\partial \mu}\right\vert_{\rm \mu=\mu_Q} = -\frac{n^2-p(\mu_Q)(\partial n/\partial \mu)_{\mu=\mu_Q}}{n^2} + 1 = 0 \Rightarrow p(\mu_Q)=0~{\rm or}~\left(\frac{\partial n}{\partial \mu}\right)_{\mu=\mu_Q} = 0 ~.
\end{equation}
While the second solution is clearly prohibited by the causality constraint (see Eq.~\eqref{eq:causality}), the first solution is not only viable but also carries the physical meaning of that ``the most stable quark matter configuration happens when the matter pressure balances the vacuum pressure''. In the right panel of Figure~\ref{fig:EOS-quark}, we show the $p=0$ line in terms of a red dashed line and mark its intersections with the thermodynamic bounds at $\mu=\mu_{Q,{\rm min,max}}(B)$ with two green crosses, which constrain the quark matter ground state energy per baryon ($E_Q(B)$) between
\begin{equation}
    \mu_{Q,{\rm min}}(B) \leq E_Q(B) \leq \mu_{Q,{\rm max}}(B) ~.
\end{equation}
This gives the most general constraint on the quark matter ground state energy per baryon that encompasses all physically possible microscopic descriptions of the deconfined QCD theory at finite baryon chemical potential.

\end{appendix}
\setlength{\bibsep}{3pt}
\providecommand{\href}[2]{#2}\begingroup\raggedright\endgroup

\end{document}